\documentclass[conference]{IEEEtran}
\IEEEoverridecommandlockouts
\usepackage{cite}
\usepackage{amsmath,amssymb,amsfonts}
\usepackage{algpseudocode}
\usepackage{graphicx}
\usepackage{textcomp}
\usepackage{makecell}
\usepackage[table]{xcolor}
\usepackage{hyperref}
\usepackage[inline]{enumitem}
\usepackage{csvsimple}
\usepackage{booktabs}
\usepackage{datatool}
\usepackage{array}
\usepackage{multirow}
\usepackage{lipsum}
\usepackage{afterpage}
\usepackage{tikz}
\usetikzlibrary{automata,arrows,positioning,calc}
\usepackage{caption}
\usepackage{float}

\font\titlefont=ptmb7t at 35pt
\font\subtitlefont=ptmr7t at 21pt
\font\tinyfont=ptmr7t at 14pt

\def\BibTeX{{\rm B\kern-.05em{\sc i\kern-.025em b}\kern-.08em
    T\kern-.1667em\lower.7ex\hbox{E}\kern-.125emX}}

\usepackage{varwidth}
\DeclareCaptionFormat{myformat}{%
  \begin{varwidth}{\linewidth}%
    \centering
    #1#2#3%
  \end{varwidth}%
  }

\title{\titlefont{ Web3Recommend }\\ \subtitlefont{ Decentralised recommendations with trust and relevance} \\ \tinyfont{ --- MSc. Thesis ---}}
\author{
\IEEEauthorblockN{Rohan Madhwal}
\IEEEauthorblockA{
Delft University of Technology\\
Delft, The Netherlands \\
R.Madhwal@student.tudelft.nl}
\and
\IEEEauthorblockN{Johan Pouwelse}
\IEEEauthorblockA{
Delft University of Technology\\
Delft, The Netherlands \\
J.A.Pouwelse@tudelft.nl}
}

\begin{document}

\maketitle
\thispagestyle{plain}
\pagestyle{plain}

\begin{abstract}

Web3Recommend is a decentralized Social Recommender System implementation that enables Web3 Platforms on Android to generate recommendations that balance trust and relevance. Generating recommendations in decentralized networks is a non-trivial problem because these networks lack a global perspective due to the absence of a central authority. Further, decentralized networks are prone to Sybil Attacks in which a single malicious user can generate multiple fake or ``Sybil" identities. Web3Recommend relies on a novel graph-based content recommendation design inspired by GraphJet, a recommendation system used in Twitter enhanced with MeritRank, a decentralized reputation scheme that provides Sybil-resistance to the system. By adding MeritRank's decay parameters to the vanilla Social Recommender Systems’ personalized SALSA graph algorithm, we can provide theoretical guarantees against Sybil Attacks in the generated recommendations. Similar to GraphJet, we focus on generating real-time recommendations by only acting on recent interactions in the social network, allowing us to cater temporally contextual recommendations while keeping a tight bound on the memory usage in resource-constrained devices, allowing for a seamless user experience. As a proof-of-concept, we integrate our system with MusicDAO, an open-source Web3 music-sharing platform, to generate personalized, real-time recommendations. Thus, we provide the first Sybil-resistant Social Recommender System, allowing real-time recommendations beyond classic user-based collaborative filtering. The system is also rigorously tested with extensive unit and integration tests. Further, our experiments demonstrate the trust-relevance balance of recommendations against multiple adversarial strategies in a test network generated using data from real music platforms. \\
\textit{Author's note: The source code, tests, and experiments performed herein are open-source and can be found on GitHub\footnote{github.com/rmadhwal/trustchain-superapp/tree/TrustedRecommendations}}
\end{abstract}


\section{Introduction} \label{intro}

The recent decade has witnessed an explosion of user-generated data on the Internet. According to a study from IBM called ``The Big Data Problem", users generate 2.5 quintillion bytes of data daily. In fact, 90 percent of the data in the world today was created in the last two years alone. \cite{lu2014toward} 

While this explosive growth in the amount of digital information available online provides a plethora of options for a diverse range of users and interests, it also results in the hindrance of timely access to items of interest and relevance since finding anything useful requires time-intensive sifting through troves of data, a majority of which is often entirely irrelevant. \cite{isinkaye2015recommendation, shenk1999data} In the words of neuroscientist Daniel J. Levitin, ``The information age is drowning us with an unprecedented deluge of data”. \cite{levitin2014organized}

The harms of information overload exceed beyond simple time wastage with studies showing that it can lead to a decrease in efficiency, increased stress, and even ill-health. \cite{bawden2020information}

The problem is exacerbated in social media platforms in the modern age, where anyone can be a content creator. \cite{deathByIO} Statistics from the popular social media platform TikTok which boasts over 1 billion monthly active users show that 83\% of the platform's users have published a video. \cite{tiktokstats}

The cardinal objective of a social media platform that aims to be successful and vibrant is an active and engaged user base. Achieving user engagement boils down to presenting the most engaging and relevant content to each user. However, popularity and success are a double-edged sword since the abundance of users and content on these platforms floods users with huge amounts of information, posing a great challenge regarding information overload. While search capabilities alleviate the problem, users often need help to express keywords that convey requirements about the type of content they would be interested in. Further, users tend to have diverse tastes, and the quality of content may be subjective depending on the user searching for it. Therefore, beyond simple searching capabilities, personalization is also required to make the content attractive and relevant to each user. \cite{gupta2013wtf, das2017survey}

A \textit{Social Recommender System} is an intelligent system that filters the massive amounts of information on social media platforms and recommends useful items and information to users based on their personalized needs, which are inferred through unique explicit and implicit interactions within the social network. In this way, Social Networks and their Recommender Systems tend to have a symbiotic relationship since the quality of recommendations catered to users allows the networks to grow, providing more interactions and higher-quality recommendations. \cite{guy2011social}

GraphJet \cite{sharma2016graphjet} is an example of a graph-based Social Recommender System used to generate Twitter content recommendations. GraphJet can provide personalized, real-time content recommendations for Twitter users, i.e., for a given user, it can recommend tweets that the user may be interested in based on the user's history and social interactions. To serve these recommendations, a personalized SALSA algorithm \cite{lempel2001salsa} is run on a bi-partite graph of interactions between users and tweets. The system assumes that a single server can store the entire graph in its memory.

However, existing Social Recommender Systems such as GraphJet are designed to work in traditional centralized social networks. In these centralized networks, users trust third-party service providers (such as Meta/Twitter/Google) with their data and provide them with recommendations. Recently, there has been an erosion in this trust due to user privacy violations, whether intentional violations through the sale of data to third parties \cite{chaabane2014closer} or unintentional violations through the loss of data through hacking breaches in the platforms. \cite{hassan2017replication}

This erosion of trust has led to a rise in the popularity of ``Web3", which leverages decentralized technologies such as distributed ledgers to offer decentralized alternatives to centralized platforms. \cite{bambacht2022web3} Web3 platforms allow direct interactions between users without third-party intermediaries or centralized servers by leveraging communal infrastructure and resources provided by the participating individuals. 

This paper addresses the challenges of generating recommendations in Web3 platforms by presenting Web3Recommend, a novel distributed Social Recommender System. Our approach integrates a graph-based content recommendation algorithm inspired by GraphJet with MeritRank \cite{nasrulin2022meritrank}, a Sybil tolerant feedback aggregation mechanism. By leveraging a personalized SALSA algorithm enhanced with Sybil-resistant random walks, we aim to balance trustworthiness and relevance in recommendations.

Web3Recommend contributes to the existing body of research by providing an end-to-end implementation that can be seamlessly integrated into any Web3 platform running on Android. We fully implemented the recommender to generate music recommendations for users of MusicDAO \cite{wissel2021fairness}, an open-source Web3 music-sharing platform that offers a decentralized alternative to Spotify/Apple Music. 

First, we describe the problem that our system aims to solve in section \ref{pd}. Next, we present the key features of our solution in section \ref{features}. Then we provide some background on the concepts discussed and techniques used in our solution in sections \ref{bgt}, \ref{bgd}, and \ref{related work}. Our system's model, assumptions, and limitations are discussed in section \ref{smaa}. In section \ref{systemdesign} we detail the implementation of the significant components in Web3Recommend. The details on how we combined the two systems to generate our recommendations are presented in section \ref{score}.

Finally, in section \ref{experiments}, we demonstrate the trust-relevance balance of recommendations through four sets of experiments. The first two sets of experiments show that the recommendations generated are relevant and that the relevance of non-Sybil nodes is not too greatly diminished with increasing MeritRank decay parameters. The last two sets of experiments involve adversarial Sybil attacks, which allow us to demonstrate the Sybil resistance of the social recommender system.

\begin{figure}[H]
    \centering
    \captionsetup{format=myformat}
    \includegraphics[height=0.8\textwidth]{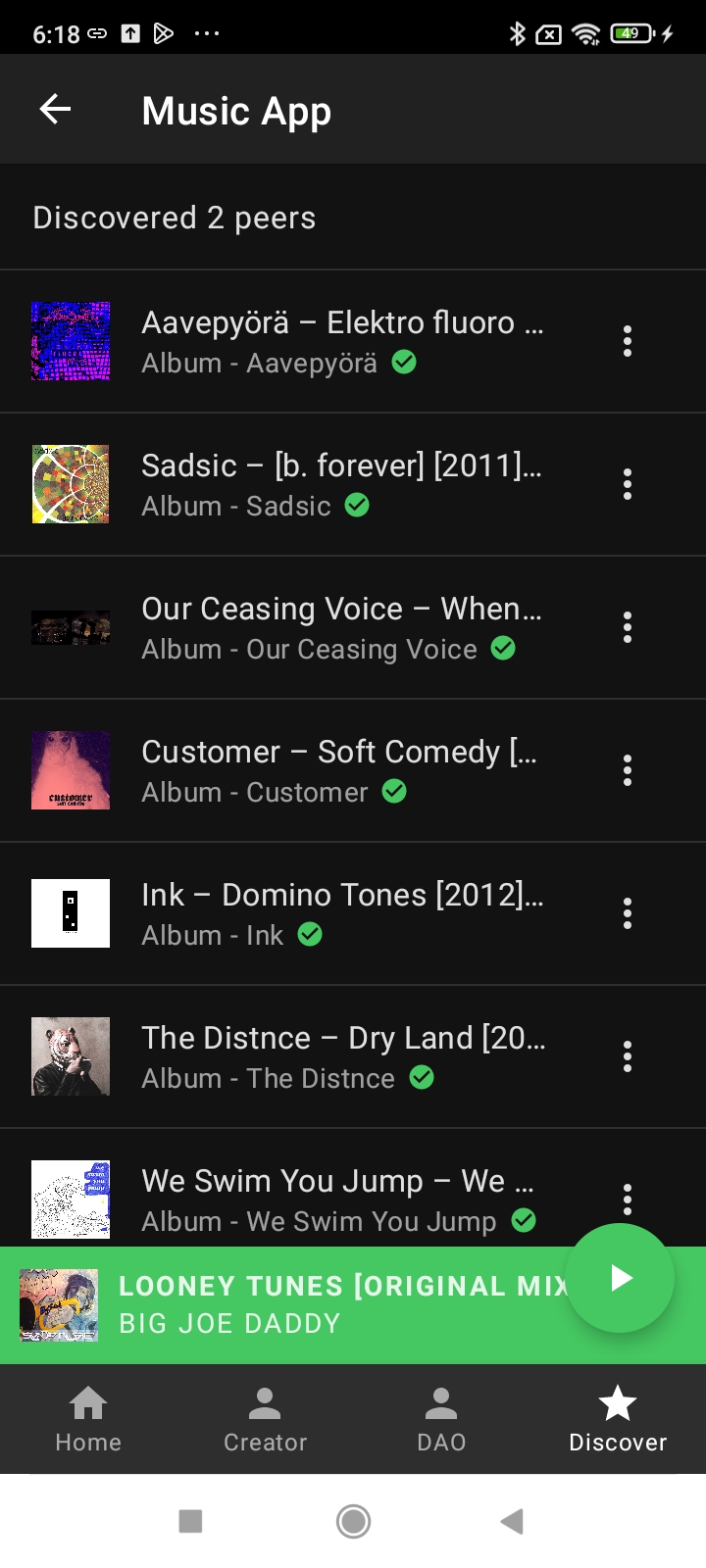}
    \caption{Recommendations generated in MusicDAO}
    \label{fig:musicdao}
\end{figure}

\section{Problem Description} \label{pd}
Generating recommendations in Web3 platforms poses unique challenges compared to traditional centralized models. Two of the main challenges are discussed below:

\subsubsection{Lack of a global perspective}

Decentralized systems require more design, planning, and management than their centralized counterparts. The main technical problem that decentralized systems face is their lack of global (centralized) knowledge. In any single node inside the system, it is difficult to have a holistic view about the rest of the network. \cite{gray1986approach}

This problem is exacerbated by the lack of a single leader who can make ad-hoc decisions in the networks, requiring all rules of the system to be pre-determined before the system goes live, further complicating the design space.  

\subsubsection{The Sybil Attack}

In conventional Social Recommender Systems, a user's prior experience can be viewed as a vote in favor of certain items. Using a Sybil Attack \cite{douceur2002sybil}, an attacker can create a potentially unlimited number of fake identities (or Sybils) to cast misleading votes. 

Thus, by creating malicious users or identities, a bad actor could mislead the system into recommending their desired items. The problem becomes important when considering the prevalence of abuse, fraud, and spam on online social media platforms. \cite{apte2019frauds} Carrying out such an attack is not incredibly complicated, with existing solutions such as Tube Automator making such attacks readily available to malicious users. \cite{Contributor_2007}

Recommender systems in centralized networks are relatively more secure from Sybil attacks due to the centralized nature of the system. Centralized systems often require user authentication and verification, making it more difficult for malicious actors to create multiple fake accounts to manipulate the ranking of items. Additionally, centralized systems have the advantage of monitoring user behavior and detecting anomalies, such as unusual activity patterns or highly repetitive actions, which could indicate the presence of a Sybil attack. This detection is possible because of the many skilled attendants dedicated to maintaining and improving system capabilities in centralized systems. These attendants can develop measures to prevent Sybil attacks, such as restricting the number of user actions that can be performed within a specific time frame or introducing identity verification requirements.

Further, the decentralized nature of Web3 platforms, coupled with their pseudonymity and anonymity, creates an environment where Sybil attacks can occur more easily. Unlike centralized systems, where user identities are typically verified, Web3 platforms prioritize user privacy and allow users to operate under multiple pseudonyms, making establishing users' authenticity and credibility challenging.

Therefore, creating a trustable and reliable decentralized recommender system for Web3 platforms is challenging due to the lack of centralized infrastructure, making it easier for malicious users to create and control multiple identities, manipulate the ranking of items, and compromise the trustworthiness of recommendations. Therefore, creating decentralized recommender systems requires new approaches that can address the challenges of decentralized networks, including Sybil attacks, limited resources, and lack of trust among users. \cite{yu2009dsybil}

\section{Key Features of Web3Recommend} \label{features}

\begin{enumerate}
    \item \textbf{Based on Monte Carlo-type methods relying on random walks} \\
    Web3Recommend uses personalized ego-centric random walks to perform computations of estimated Personalized Page Rank and SALSA values for nodes in the network. These values are then used to generate recommendations in the system. It has been shown that Monte Carlo methods can provide very good probabilistic estimations for Page Rank and SALSA. They are also much faster and parallelizable than the conventional power iteration method, making them a good choice for an online recommendation system. \cite{avrachenkov2007monte} Additionally, by enhancing the random walks in these methods with decay parameters from MeritRank, Web3Recommend is also able to limit the influence of Sybils in these estimations. They also allow us to create a simple, understandable, yet sufficiently expressive system to generate relevant, trustworthy recommendations. Random walks allow us to define a large design space, allowing room for customization to various use cases in different applications (e.g., social search) and contexts. Further, random walks act as ``social proof" for the recommendations, allowing users to understand better why certain items were recommended, leading to higher user engagement. There is also room for further increasing the quality of generated recommendations by feeding the output from our random walks as input to machine learning models. Still, in our case, the direct output is sufficient for user consumption.
    \item \textbf{Each node stores the entire interaction graph between users and items, which is synchronized using an edge gossiping mechanism} \\
    While readers might find storing the entire graph in a single node/device strange, graph partitioning remains a complex problem in large, dynamically changing graphs despite much work and progress in the field. \cite{leskovec2008community} Achieving graph partitioning would require implementing a fully-distributed graph progressing engine and further add high communication costs to the system. Additionally, in a P2P Web3 platform, users are not expected to be always available/online. Thus, like GraphJet, Web3Recommend assumes that all nodes store the entire graph in memory. Our compact graph serialization techniques enable storing up to a billion edges in less than 8GB of memory. Given the hardware present in modern devices and Moore's Law, this assumption is not very unreasonable.
    \item \textbf{Bootstrapping mechanisms} \\
    The ``new user" problem in Social Recommender systems \cite{rashid2002getting} necessitates bootstrapping mechanisms to introduce new users to the network. Web3Recommend includes two bootstrapping mechanisms: 1) A similarity-based mechanism for finding similar users, allowing new users to find users they can trust 2) A personalized page rank for creating a circle of trust which can be used for recommending relevant items to users who haven't consumed many items and thus don't have any existing edges in the interaction graph
\end{enumerate}

\section{Background on Trust} \label{bgt}

The system presented in this paper relies on the (incremental) computation of personalized PageRank and SALSA augmented with principles from MeritRank. We also build on top of the GraphJet recommender system by Twitter. In this section, we provide a quick review of these methods.

\subsection{PageRank}
One of the world's most widely known ranking systems is Google’s PageRank \cite{page1999pagerank}, which is still used (along with other algorithms) to rank websites for user queries on Google. 

PageRank determines a rough estimate of the relative importance of a website by computing a ranking for every web page. The underlying assumption of PageRank is that a more important website is more likely to receive links from other websites than a less important website i.e., that the existence of a hyperlink $u \rightarrow v$ implies that the page $u$ votes for the quality of page $v$ and hence, the most important page receives most votes. PageRank gave birth to topic-sensitive or personalized ranking and other hyperlink-based centrality measures. \cite{10.1145/371920.372096} 

More formally, let $V$ represent the set of all web pages in a network. A \textit{web graph} is the directed graph which consists of the vertex set $V$, and the hyperlinks between pages represent the edges in the network. Further, let $r$ be the \textit{preference vector} which induces a probability distribution over $V$, and $c \in (0,1)$ be the \textit{reset probability}.

Then, the PageRank vector $p$ is the solution of the following equation: \cite{fogaras2005towards}

\begin{equation}
    p = (1 - c) \times pA + c \times r
\end{equation}

If $r$ is uniform over $V$ then $p$ is referred to as the \textit{global PageRank vector}. In this paper, the special case where for some web page $x$, the $x^{th}$ coordinate of $r$ is 1, and the rest of the coordinates are 0, the solution of $r$ represents the \textit{Personalized PageRank} of web page $x$ and is denoted as PPR($x$).

Note that it is also equivalent to interpret PageRank with the Random Surfer model \cite{blum2006random} where PageRank is simply the stationary distribution of a random walk where at each step, assuming we are at a specific web page $u$, with probability $c$ we jump to a random web page $v$, and with probability $1 - c$ we follow a randomly chosen outgoing edge (or hyperlink) from $u$ to a new web page $w$. In this model, Personalized Page Rank is the same as PageRank except that all random walks start and jump randomly to the seed node $x$ for which we are personalizing the PageRanks. \cite{bahmani2010fast}

\subsection{SALSA} \label{salsa}

Stochastic Approach for Link-Structure Analysis or SALSA is a web page ranking algorithm similar to PageRank and HITS \cite{kleinberg1999authoritative}, which attempts to extract information from the link structure of a networked environment to associate two scores with every node $v$, the \textbf{hub score} $h_v$ and \textbf{authority score} $a_x$. As the name suggests, $a_x$ reflects how much of an \textit{authority} a node is on a specific topic. While the notion of authority is pretty broad, the intuitive idea is that in the context of a particular query, links from one node to another express a considerable amount of latent human judgment, and that judgment is precisely what is needed to formulate the notion of authority. Hence, if many nodes point to another node, it will possess a high authority score. $h_v$, on the other hand, reflects how well nodes point to authorities, so if a node largely links to other nodes that are considered very authoritative, it will have a high  $h_v$.

More formally, if $E$ is the set of all edges in the graph and $indeg(x)$ and $outdeg(v)$ are the in-degrees and out-degrees of the node respectively then:

\begin{equation}
    h_v = \sum_{\{x | (v,x) \in E\}} \frac{a_x}{indeg(x)}
\end{equation}

\begin{equation}
    a_x = \sum_{\{v | (v,x) \in E\}} \frac{h_v}{outdeg(v)}
\end{equation}

It is worth noting that, unlike PageRank, where we only have forward random walks, SALSA consists of a forward-backward random walk where the walk alternates between forward and backward steps. 

Similar to Personalized PageRanks, we also have Personalized SALSA, which tailors the hub and authority scores to a single root node. As in Personalized PageRank, we can have random jumps to the seed node in the personalized version of SALSA. Assuming that the seed node is $u$, $h_{v,u}$, the personalized SALSA hub score and $a_{x,u}$, the personalized SALSA authority score can be represented as:

\begin{equation}
    h_{v,u} = c \times \delta_{u,v} + (1 - c) \times \sum_{\{x | (v,x) \in E\}} \frac{a_x}{indeg(x)}
\end{equation}

\begin{equation}
    a_{x,u} = \sum_{\{v | (v,x) \in E\}} \frac{h_v}{outdeg(v)}
\end{equation}

Notice that in this setting, the hub and authority scores can be interpreted as the similarity and relevance scores, respectively. This ability of personalized SALSA to create tailored recommendations for specific root nodes is used later on inside our recommendation system.

\subsection{GraphJet} \label{graphjet}

\begin{figure}[t]
    \centering
    \captionsetup{format=myformat}
    \includegraphics[scale=0.5]{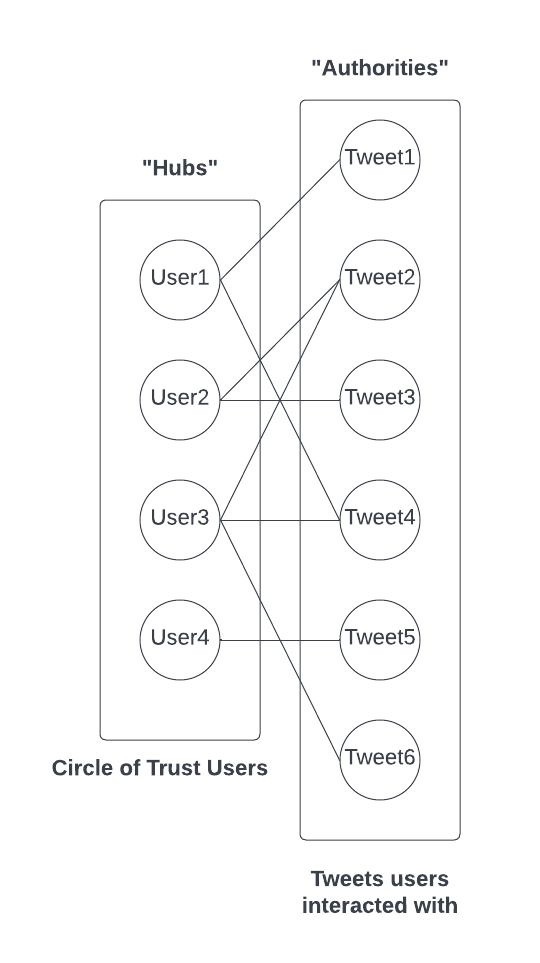}
    \caption{Example of bi-partite graph used in GraphJet}
    \label{fig:graphjetgraph}
\end{figure}

GraphJet is a graph-based system for generating real-time tweet recommendations on Twitter. The recommendation algorithm is based on an adaptation of personalized SALSA, which involves random walks in a bi-partite graph of users and tweets. Formally, GraphJet manages a dynamic, sparse, undirected bipartite graph $G = (U,T,E)$ where $U$ represents users in Twitter, $T$ represents tweets, and $E$ represents interactions between the users and tweets over a temporal window. Hence, the bipartite graph $G$ consists of two sets of nodes $U$ and $T$. Users from $U$ are always on the left side of the random walk and represent \textit{hubs}, while tweets from $T$ are on the right side and represent \textit{authorities}. Figure \ref{fig:graphjetgraph} demonstrates a sample bi-partite graph used by GraphJet, where the left side consists of users from the ``circle of trust" of the user whose recommendations are being generated, and the right side includes tweets the users in the circle interacted with. The circle of trust is constructed using a personalized PageRank algorithm.

GraphJet maintains and updates the bipartite graph by keeping track of user-tweet interactions over the most recent \textit{n} hours. Periodically, edges older than \textit{n} hours are discarded to ensure that memory consumption doesn't increase boundlessly. Experiments show that this pruning does not have a noticeable impact on the system's recommendation quality. The system supports high-performance ingestion of real-time interactions and generations of recommendations. 

Below is a simplified description of the SALSA algorithm run inside GraphJet:

\begin{enumerate}
    \item The random walk begins from the vertex $u$ in the left-hand side of the bi-partite graph corresponding to the querying user
    \item An incident node from $u$ to the right-hand side of the graph is uniformly selected to a tweet $t$ on the right-hand side of the bi-partite graph
    \item From $t$, an incident edge is selected uniformly back to the left-hand side to another node $v$
    \item This is repeated an odd number of steps
\end{enumerate}

\begin{figure}[t]
    \centering
    \captionsetup{format=myformat}
    \includegraphics[scale=0.25]{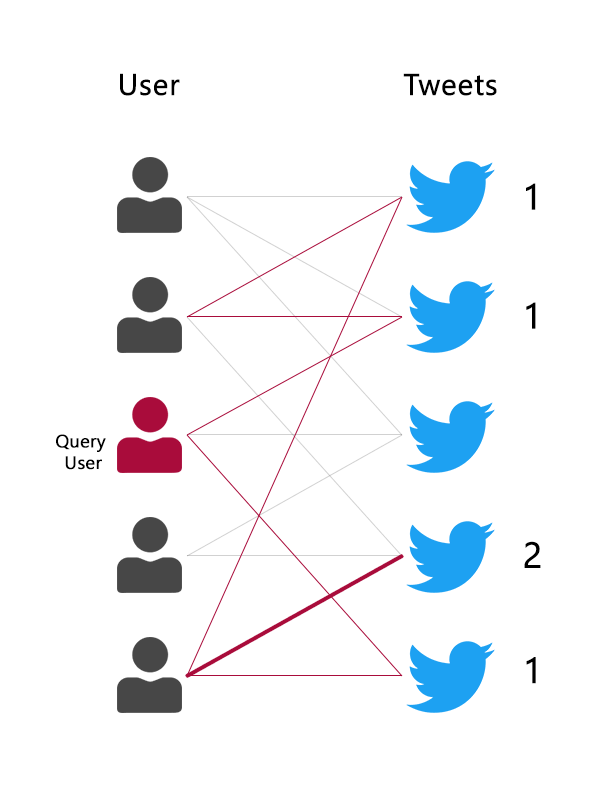}
    \caption{Example of a random walk in GraphJet's Personalized SALSA}
    \label{fig:graphjetrandomwalk}
\end{figure}

Figure \ref{fig:graphjetrandomwalk} demonstrates a sample random walk in one iteration of the above SALSA algorithm. 
To introduce personalization, a reset probability as described in \ref{salsa} above is used, which restarts the random walk from vertex $u$ to ensure that the random walk doesn't ``stray" too far from the query vertex. 
Further, it may also be possible that the querying user doesn't have any existing interactions in the bi-partite graph, either because the last interaction was more than \textit{n} hours ago or because they are a new user. In this case, the random walk could start from a \textit{seed set} instead of from the query user's node. This seed set is configurable, but the usual choice is to use the user's circle of trust constructed using Personalized PageRank.

After constructing the bi-partite graph, multiple instances of the SALSA algorithm are run, assigning hub scores to the left side and authority scores to the right. The vertices on the right-hand side are then ranked and presented as tweet recommendations to the user. The vertices on the left-hand side are also ranked and, based on the homophily principle, can be additionally catered as ``similar user" recommendations.

The GraphJet paper suggests that the algorithm is effective because it is able to capture the recursive nature of the user recommendation problem. A user $u$ is also bound to like tweets liked by users similar to u. These users are in turn similar
to u if they follow the same (or similar) users. Personalized SALSA operationalizes this idea, providing similar users to $u$ on the left-hand side and tweets they like on the right-hand side. The random walk in SALSA also ensures equitable distribution of scores out of the vertices in both directions.

\section{Background on Decentralisation} \label{bgd}

\subsection{Decentralisation}

\begin{figure*}[t]
    \centering
    \captionsetup{format=myformat}
    \includegraphics[width=0.7\textwidth]{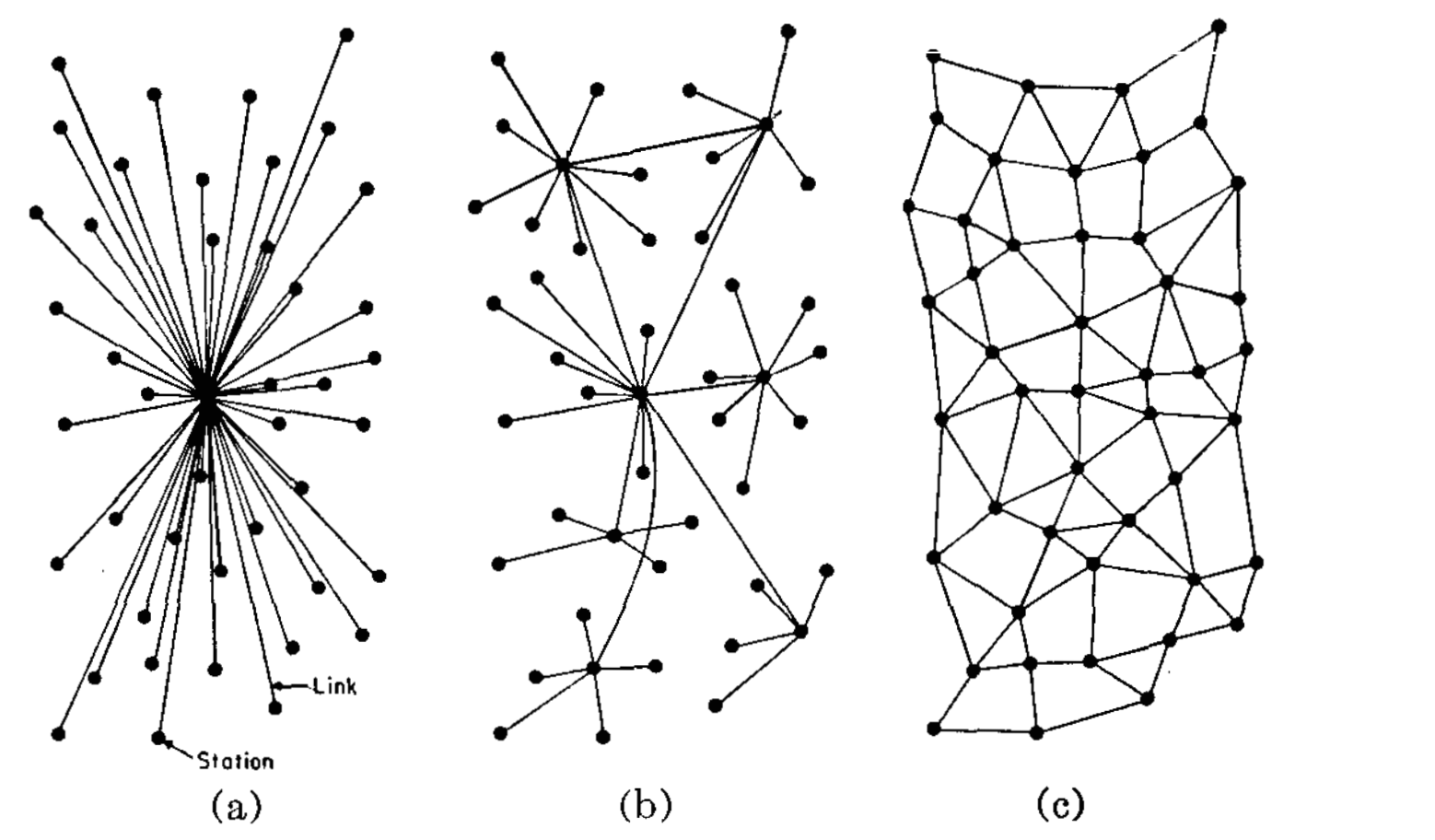}
    \caption{a) Centralized network b) Decentralised network c) Distributed network\newline
    Cardinal architectural insight from Baran's 1964 paper \cite{baran}}
    \label{fig:decen}
\end{figure*}

The term ``decentralized network" was initially introduced by Paul Baran, one of the pioneers of packet switching. Networks can generally be classified into two types: ``star" or centralized networks, and ``grid" or distributed networks. In a star/centralized network, all nodes are connected to a central node, requiring participants to go through this central component to interact with one another. In contrast, distributed networks have no central node, enabling direct communication between nodes without reliance on a centralized point. 

Baran termed networks that utilized a combination of centralized and distributed components as ``decentralized" since they lacked a single, central point of failure. \cite{baran} Figure \ref{fig:decen} illustrates these differerent network types.

In contemporary literature, the term ``decentralized network" refers to networks where instead of relying on large centralised platforms, participants and contributors control the technology, content, and infrastructure. This control manifests in various ways, such as participants managing specific parts of the infrastructure, collaborators owning their own private data silos that are queried during network discovery, and participants having autonomy in determining the operational details of the network. \cite{Korpal2022} 

Twitter serves as an example of a centralized network, where the platform owns all user-generated content. Conversely, Tribler, a peer-to-peer file-sharing system that builds upon the BitTorrent protocol, exemplifies a decentralized network. Tribler enables users to discover content using keyword searches and incorporates a reputation-management system to foster collaboration. \cite{tribler}

\subsection{Web3}
The term ``Web2.0" was initially introduced by Tim O'Reilly in 2007 to describe a new iteration of the Internet that empowered users to publish, consume, and interact with content and each other. \cite{o2007web} It aimed to expand upon the earlier version, ``Web1.0", which primarily featured static pages for information display. While ``Web1.0" was often referred to as the ``read web", ``Web2.0" aimed to be the ``read-write web".

Critics, including Tim Berners-Lee, the inventor of the World Wide Web, argue that ``Web2.0" failed to fulfill the vision of a secure, decentralized exchange of public and private data. Instead, users' data became increasingly stored in corporate data silos, raising concerns about data ownership and security. Berners-Lee and others advocate for users to own their data to ensure data security. \cite{solid}

In 2014, Gavin Wood, co-founder of Polkadot and Ethereum, introduced the term ``Web3.0" to describe an Internet that is decentralized, open, and transparent. The Web3 movement seeks to transform the platform-oriented ``Web2.0" into a decentralized web ecosystem with the following goals: 1) Avoiding content discovery and propagation monopolies by large centralized entities, 2) Supporting immersive web development, 3) Combating the spread of misinformation, and 4) Enabling platform users to create, exchange, and react to information securely, privately, and freely. \cite{wood}

\subsection{EigenTrust} \label{eigentrust}

Establishing trust among peers in decentralized networks is crucial for secure and reliable interactions. Trust mechanisms play a vital role in determining the credibility of peers and ensuring that transactions occur only with trustworthy counterparts. 

EigenTrust \cite{eigenTrust} was one of the pioneering methods to address this trust challenge in decentralized networks. EigenTrust provides a distributed algorithm that calculates a global trust value for each peer, reflecting the collective experience of all peers in the network, enabling peers to make informed decisions about the trustworthiness of their counterparts, thereby fostering trustworthy interactions. In this section, we discuss the workings of EigenTrust and its significance in decentralized network environments.

The primary objective of the algorithm is to determine a unique global trust value, denoted as $\overrightarrow{t}$, for each peer in the network. This value encapsulates the overall experience of all peers with a given peer and serves as a reference for evaluating the trustworthiness of other peers. Furthermore, EigenTrust incorporates mechanisms to prevent malicious groups of collaborating peers from providing deceptive trust ratings for their advantage.

Similar to eBay's reputation system \cite{kollock1999production}, EigenTrust requires each peer to rate its transactions with other peers, thereby generating a local trust value, denoted as $s$, for each peer. The value $s_{ij}$ represents the trust level of peer $i$ towards peer $j$, based on their previous transactions. A suggested approach for calculating $s_{ij}$ is through the following formula:

\begin{equation}
s_{ij} = sat(i,j) - unsat(i,j)
\end{equation}

Here, $sat(i,j)$ and $unsat(i,j)$ represent the number of satisfactory and unsatisfactory transactions that peer $i$ has conducted with peer $j$. These localized trust values, $s_{ij}$, are normalized to obtain $c_{ij}$, ensuring that trust values fall between 0 and 1. This normalization prevents malicious peers from arbitrarily assigning extremely high trust values to other malicious peers while assigning low values to trustworthy peers, thereby exploiting the system.

To derive the global trust value, $t$, these local trust values are aggregated within each peer using the equation:

\begin{equation}
t_{ik} = \sum_j c_{ij} c_{jk}
\end{equation}

This calculation can be interpreted as peer $i$ seeking opinions from its acquaintances regarding how much they trust their acquaintances. However, this process needs to be iterated to reflect the experiences of acquaintances' acquaintances. The authors of the EigenTrust paper demonstrate that the final trust vector, denoted as $\overrightarrow{t_i}$, will converge to the same vector $\overrightarrow{t}$ for every peer $i$ in the network, representing the left principal eigenvector of the matrix $[c_{ij}]$.

EigenTrust calculates $\overrightarrow{t}$ in a distributed manner. The authors establish that the algorithm can be executed relatively efficiently in networks with a small number of active peers, as each peer engages in a limited number of transactions.

EigenTrust addresses the primary challenge of designing distributed reputation systems—aggregating local trust values into global trust values and by doing so, introduced the concept of ``transitive trust".

Although EigenTrust presents a robust method for establishing trust in decentralized networks, it necessitates an initial notion of trust, which typically comprises a group of known trustworthy peers. The authors suggest that this initial group could consist of the early adopters and designers of a peer-to-peer network, as they are less likely to engage in fraudulent behavior within a network they helped create.

By employing EigenTrust or similar trust mechanisms, decentralized networks can establish a foundation of trust among their peers, enabling secure, reliable, and transparent interactions that are not reliant on centralized authorities.

However, while EigenTrust allows the calculation of trust in other users, the transitive trust calculated by it is not Sybil-resistant i.e., Sybil attacks can be used to increase the perceived trust in Sybil identities inside the system.

\begin{figure*}[t]
    \centering
    \captionsetup{format=myformat}
    \includegraphics[width=0.9\textwidth]{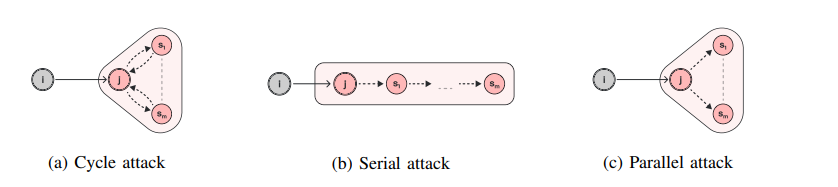}
    \caption{Sybil attack strategies \cite{meritRank}}
    \label{fig:sybil_attacks}
\end{figure*}

\subsection{MeritRank} \label{meritrank}

\textit{MeritRank} \cite{meritRank} aims to bound the benefits of Sybil attacks instead of preventing them altogether. The system is based on the assumption that peers observe and evaluate each others' contributions. Each peer's evaluation is stored in a personal ledger and modeled in a feedback graph where the feedback to each user is modeled as a special token value accumulating over time. It is also assumed that each peer can discover the feedback graph, for example, through a gossip protocol.

To limit the influence of Sybils, three main types of Sybil attack strategies are identified in MeritRank. These are illustrated in Fig. \ref{fig:sybil_attacks}.

MeritRank manages to achieve Sybil tolerance by imposing the following constraints on how reputation can be gained inside the feedback graph when using any of the above strategies:
\begin{enumerate}
    \item \textbf{Transitivity $\alpha$ decay} \\ This constraint limits the ability of an entity to create a serial Sybil attack by terminating random walks in the feedback graph with a probability $\alpha$
    \item \textbf{Connectivity $\beta$ decay} \\ Sybil attack edges in a feedback graph are often bridges i.e., their cut creates two separate components. This constraint introduces punishment for a node for being in a separate component
\end{enumerate}
A trust graph modeled using these MeritRank's constraints will satisfy:
\begin{equation}
\lim_{|S|\to\infty} \frac{w^+(\sigma_s)}{w^-(\sigma_s)} \leq c \\
\end{equation}
where, $w^+(\sigma_s)$ is the profit gained by the Sybil Attack $\sigma_s$, $w^-(\sigma_s)$ is the cost of the Sybil attack, $S$ is the set of Sybils and $c$ is some constant value such that $c > 0$. Thus, MeritRank can provide a reputation system with feedback that is Sybil tolerant.

\section{Related Work} \label{related work}
In this section, we cover broad categories of existing work in research which attempt to solve similar problems and show why they do not achieve the goals that we try to fulfill. Thus, we demonstrate the relevancy of our work by pointing out how it serves the interesting niche of generating trustworthy and relevant recommendations in Web3 platforms.
\subsection{Bounding Identity Creation against Sybils}
A popular method to defend against Sybil attacks is to leverage defenses that bound the ability of malicious attackers to create Sybil identities and hence indirectly limit the influence of the Sybil attack by limiting the votes that a Sybil attack can cast.
The most rudimentary method of achieving this is ensuring that all unique identities on platforms correspond to real human beings. This can be achieved using a trusted central authority to verify information about users unique to human beings, such as passports, phone numbers, credit cards, etc. Other approaches involve using graphical challenges such as CAPTCHA to ensure the user is a human. \cite{borisov2006computational} Such approaches are  flawed in the context of Web3 for multiple reasons. First, users of Web3 platforms employing such methods may be hindered from using them because of privacy concerns. The anonymity guarantees of these platforms are often the primary reason many users choose to use these platforms. Second, using a centralized third party for verification is antithetical to the idea of decentralization that Web3 platforms stand for, and further, maintaining a centralized server adds too much complexity to the system's design.

Decentralized approaches to the same problem have been suggested in research, such as limited identities from specific IP addresses/prefixes \cite{damiani2002reputation}, creating resource-based challenges \cite{rowaihy2007limiting} and remote issuing of certificates to verify identity based on network/location coordinates  \cite{bazzi2005establishment}.
While these approaches make it harder to create Sybil identities, they do not entirely stop Sybil attacks, and a powerful adversary with enough incentive could easily surpass these mechanisms to generate significant influence on the recommendations in the network. Further, bounding identity creation can also have the unintended effect of making it frustrating for non-Sybil users to use the platform. Therefore, our work does not rely on bounding user creation to limit Sybil influence indirectly. Instead, it directly limits Sybil influence by effectively detecting Sybils and restricting the influence of their votes.

\subsection{Reputation Systems}
Reputation Systems \cite{resnick2000reputation} allow the collection of feedback from different users in the network to determine which peers can be trusted based on their past behavior. A famous example of a reputation system is the ``Feedback Forum" on eBay \cite{resnick2006value} : after a transaction is completed, a buyer or seller can rate each other (1, 0 or -1) and leave comments. A participant in eBay accumulates such points over time, which are displayed next to their screen name. A buyer can view a seller’s points and comments left by other users to create a ``shadow of the future” into the transaction they can expect to have if they buy an item from the seller. Many other online forums and marketplaces, such as Amazon and Stack Overflow, rely on similar reputation systems. However, while reputation systems are a strong mechanism to determine whether certain users are trustworthy, they do not provide a way to generate recommendations based on trustworthy users.

\subsection{Social Network-based Sybil Defense}
Other works have utilized the power of feedback from social networks to limit Sybil influence. Popular examples are SybilGuard \cite{yu2006sybilguard}, SybilLimit \cite{yu2008sybillimit}, Ostra \cite{mislove2008ostra} and SumUp \cite{tran2009sybil}. Similar to our work, these methods can establish an approximate notion of trust among users using the properties of graphs and often assume global knowledge of the dynamically changing social network. Using these properties, the influence of Sybil identities in the votes is limited. However, similar to vanilla Reputation Systems, while these approaches are great for generating trust in the network, on their own, they cannot ensure that the non-Sybil items they identify are relevant to the user for whom the recommendations are being generated. We achieve this by building on personalized SALSA and ensuring that the recommended item is trustworthy (i.e., non-Sybil) yet also relevant to the user for which the recommendations are being tailored. 

\subsection{Machine Learning-based approaches}
Existing machine learning-based approaches to recommender systems attempt to apply techniques from the multiarmed bandit problem \cite{auer2002nonstochastic} or the contextual bandit problem \cite{li2010contextual} where contextual information is used to group users that belong to the same cluster via classification or clustering techniques. The problem with this approach is that they assume the presence of considerable existing feedback from users on what items they like and the ``goodness" of various objects since these models are only as good as the data they are trained on. Web3 is an emerging technology, and in many platforms (especially new platforms), this data needs to be more sizable to train decent models, leading to poor recommendations or the inability to have such models altogether. \cite{yu2009dsybil}

\subsection{Other approaches}
Another notable system worth mentioning is Dsybil \cite{yu2009dsybil}. The system presented in the paper utilizes similar mechanisms for diminishing Sybil influence and generating recommendations as our paper. They can achieve this by: i) exploiting
the heavy-tail distribution of the typical voting behavior of
the honest identities and ii) carefully identifying whether the
system is already getting ``enough help” from the voters, and hence, if Sybil votes are only latching on to existing votes. While they demonstrate an impressive Sybil tolerance, a notable drawback of their paper is its reliance on only explicit feedback through voting on items that need to be recommended. However, explicit feedback is not always available, and in fact, most of the feedback on social networks is implicit rather than explicit. \cite{guo2015learning} Our system incorporates implicit feedback from users to generate recommendations. Further, like many other mentioned papers, the system focuses purely on trust and not on generating relevant, personalized recommendations.

\section{System Model, Assumptions and Limitations} \label{smaa}

\subsection{Items, Users and Votes}
Web3Recommends recommends \textit{items} (e.g., songs/albums in Spotify, movies in Netflix, posts in Reddit, etc.) to the platform's \textit{users} based on the experience of the users with the items. A user's preference for an item serves as a \textit{vote} for the item. Therefore, if enough users like a particular item, it is likely to have more votes by virtue of being visited more in random walks inside the personalized SALSA algorithm and, therefore, more likely to be recommended to other users.

\subsection{Target Application/Scenario} \label{applications}
Recommendation Systems are a broad concept; different systems differ in their goals and details. \cite{bobadilla2013recommender} Therefore, a solution that works in one scenario may not work in another scenario because the platform requires a different purpose from its Recommendation System. For example, a Recommendation System in an online retailer such as eBay may be required to generate all products that a user may be interested in so that the user has multiple choices. In contrast, the system in modern media applications such as TikTok would only be required to generate a single recommendation to autoplay as the next item for the user.

Web3Recommend aims to cover a broad range of use cases by providing a system that can rank all items available to a particular user. For this, it assumes scenarios where: 
\begin{enumerate}
    \item The objects to be recommended can be uniquely identified and are always available to the user (the exact method of availability could vary from peer-to-peer distribution to provision by a central server)
    \item The lifespans of the users in the network are not incredibly short-lived, allowing them to establish trust relationships with other users
    \item Users can initially discover trusted users using social discovery mechanisms, or an initial set of trusted users are provided to the user by the platform
\end{enumerate}

The final assumption is a significant limitation of our work since it's impossible to build trust out of nothing, and a new user can end up trusting Sybil users via bootstrap threats. Note that this is not an uncommon assumption with EigenTrust \ref{eigentrust} imposing a similar assumption. 

While relying on a selected group of peers shares a lot of problems as relying on a centralised entity, this problem can be alleviated by dynamically calculating the trusted users in the network instead of using a static set, similar to the solution presented in HonestPeer. \cite{kurdi2015honestpeer}

\subsection{Network Assumptions}
Web3Recommend assumes that the application utilizing it leverages a peer-to-peer architecture \cite{ripeanu2001peer} with each user operating their node and possessing the ability to communicate directly with any other node and item in the network. We assume that all users in the network follow the protocol honestly and cannot tamper with it in any way. We also assume that communication occurs over privacy-preserving, cryptographically secure protocols.

While the system presented in this paper relies on Gossiping \ref{gossiping} over a content overlay network to synchronize its data, it is essential to acknowledge that it currently lacks sufficient security measures to protect against spoofing of gossip. However, it is worth exploring the potential of enhancing the system's security by implementing a certificate-based approach. By utilizing certificates for each node within the content overlay network, it becomes possible to establish a secure communication framework. These certificates can verify the authenticity and integrity of gossip messages exchanged between nodes, ensuring that only trusted and authorized nodes participate in the synchronization process. Moreover, by incorporating individual certificates for each user, the system can prevent users from pretending to be another user and signing actions on their behalf. This significantly raises the bar for spoofing attempts, enhancing the overall security of the gossiping process.

Further, we rely on timestamps to detect when edges were created, thus we assume that the clocks across different nodes are synchronized to prevent ordering issues.

\subsection{Affinity and Trust} \label{affinitytrust}
Web3Recommend uses two distinct concepts, \textit{affinity} and \textit{trust}, to model user-to-user and user-to-item relationships respectively. 

A user $u$'s affinity for item $i$ is expressed by $Af(u,i)$ and is calculated by:

\begin{equation}
    Af(u,i) = \frac{PC(u,i)}{\sum_{x \in I_u} PC(u,x)}
\end{equation}

Where $PC(u,i)$ is the \textit{play count} of user $u$ for item $i$ i.e. the number of times the user has consumed item $i$, and $I_u$ is the set of all items that have been consumed by user $u$. Therefore, $Af(u, i)$ is a value between 0 and 1 and serves as a measure of the user $u$'s preference for item $i$ compared to all the items that they have consumed.

This definition of affinity may only suit some use cases, and more fine-grained metrics, such as the ratio of the item's playtime to the total playtime could be more appropriate. 

Like in reputation systems, a user's trust in another user is increased when the user performs valuable work for the other, which in this case of our recommendation systems means providing a recommendation that the other user likes, i.e., ends up developing an affinity for. Therefore, affinity and trust are inherently linked since as a user's affinity for an item increases, the user's trust in other users who recommended the item also increases.

Since, in our system, recommendations are calculated using random walks inside a personalized SALSA algorithm, recommending an item to another user means that the recommender ``voted" for the item in the random walk by having a high affinity for it.

Note that modeling trust this way ensures that Sybil users cannot simply piggyback on popular items to increase the popularity of their Sybil items since: 1) if they have a high affinity for popular items, their Sybil items are less likely to be visited 2) On the other hand, if they have a low affinity for those popular items, the random walk is less likely to visit them 3) Many users are likely to have a high affinity for popular items thus, Sybils are less likely to benefit from following this strategy 4) To gain trust, the Sybils still need to effectively perform ``useful work" by recommending items that other users like. Hence, They cannot simply reap rewards from items that have already been recommended to other users.

\section{Web3Recommend Architecture and design} \label{systemdesign}

Web3Recommend is a Social Recommender System designed to provide recommendations for any application running on a decentralized network. The central data structure in the network is the \textbf{TrustNetwork} which stores information about user-to-user and user-to-item relationships across the entire network. 

Two random-walk based algorithms are run on top of the TrustNetwork: 1) A Personalized Page Rank calculates a global trust value for each user using the notion of ``transitive trust" as presented in EigenTrust \ref{eigentrust} 2) A Personalized SALSA algorithm that calculates a recommendation score for each item in the network

The random walks in both these algorithms are augmented with MeritRank to provide Sybil-resistance.

Each node maintains a personal copy of a TrustNetwork, and updates to the network are synchronized through a timestamp-biased edge gossiping mechanism, ensuring that recommendations are based on recent, global information inside the network. The system design also includes a simple bootstrapping mechanism that allows new users to find similar users in the network. However, it is worth noting that malicious users can exploit this bootstrap mechanism. In an actual application, we assume that users are able to bootstrap through the social discovery of trusted peers or through the provision of trustworthy nodes by the application itself. 
The following is an in-depth description of the various components of the system:

\subsection{TrustNetwork}

As mentioned before, TrustNetwork is the central data structure of Web3Recommend. TrustNetwork consists of two types of nodes: users, $U$, and items, $I$. Further, there are two types of edges in the network: directed user-to-user weighted edges representing the trust a user places in another user, and undirected user-to-item weighted edges representing the affinity of a user for an item. In practice, this is implemented using a combination of two graph structures:
\begin{enumerate}
    \item \textbf{User to User Graph} \\
    The user-to-user graph is a \textit{weighted directed acyclic graph} in which the graph's vertex set consists of all users in the network, and the edge set consists of trust relationships between the users. To ensure efficient memory usage and to guarantee that an edge between two users only exists if they trust each other, only the top 5 edges in terms of trust/weight are retained for each user.
    \item \textbf{User to Item Graph} \\
    The user-to-item graph is a \textit{weighted undirected acyclic bi-partite graph} in which the vertex set of the graph consists of all users and items in the network and the edge set consists of affinity relationships between the users and items
\end{enumerate}

All algorithms in Web3Recommend operate on top of the TrustNetwork. In our Kotlin implementation, we use JGraphT \cite{michail2020jgrapht} to efficiently implement both graph data structures.

\subsection{Recommendation Algorithm}
The main component of Web3Recommend is its elegant recommendation algorithm. The algorithm is inspired by GraphJet's algorithm presented in \ref{graphjet}. 

We present three modifications to the personalized SALSA used in the original algorithm. 

\begin{enumerate}
\item \textbf{Weighted Random Walks} \\
Instead of using uniform probabilities to decide which node to walk to in our random walks, we perform walks using the affinity metric defined earlier. Hence, when walking from a node to an item the affinity for the item biases the walk, so if a user $u$ prefers an item $i_j$ two times more than item $i_k$, the random walk is also twice more likely to visit $i_j$ from $u$ than $i_k$. We believe this is a reasonable assumption that aligns with our goal of recommending items that similar users like. Similarly, when walking back from an item to a user, the walks are biased by the user's affinity for the item. Hence, we are more likely to travel to a user with a higher affinity for the item. Again, finding similar users boils down to finding users who like the same items, so it's reasonable to bias our walks this way.

\item \textbf{Add MeritRank decays to limit the influence of Sybil attacks} \\
Alpha and beta decays from MeritRank are added to the system to add Sybil tolerance. In \ref{score}, we explain how the decays are calculated and used to provide Sybil-resistant recommendations. In our experiments, we vary these decay values to measure their influence on trust (Sybil tolerance) and the relevancy of recommendations. 

\item \textbf{Generate ``trusted" random walks} \\
While GraphJet's personalized SALSA algorithm allows the generation of personalized, relevant recommendations, the recommendations generated are not guaranteed to be trustworthy. This is because when performing a random walk from an item to a user, it's possible to walk to a nontrusted (Sybil) user who also claims to like the item. In Web3Recommend, we modify the walk back from an item to a user to be limited to users who the original voter of the item trusts. This ensures that if we reach a Sybil user in a random walk, it's through a malicious user or another Sybil user. Hence, their influence on voting can be detected and limited through MeritRank's decays.
\end{enumerate}

A simplified version of a random walk in our personalized SALSA algorithm is illustrated in \ref{alg:salsa}.
\newcommand{\MiniWrp}{\par\qquad\enspace}
\begin{figure*}[t]
\begin{algorithmic}[1]
\Function{RandomWalk}{$currentNode$, $lastNode$, $userToUserGraph$, $userToItemGraph$}
    \If{currentNode is a User}
        \State WalkToSong($currentNode$, $lastNode$, $userToItemGraph$)
    \Else
        \State WalkToUser($currentNode$, $lastNode$, $userToUserGraph$, $userToItemGraph$)
    \EndIf
\EndFunction
\item[]
\Function{WalkToSong}{$currentNode$, $lastNode$, $userToItemGraph$}
    \State $userEdgesWeight \gets 0$
    \ForAll{edge $e \in$ \{edges of $currentNode$\}}
            \State $userEdgesWeight \gets userEdgesWeight + userToItemGraph.WeightOf(e)$
    \EndFor
    \State $p \gets userEdgesWeight * RandomDoubleBetween(0, 1)$
    \State $cumulativeP \gets 0$
    \ForAll{edge $e \in$ \{edges of $currentNode$\}}
            \State $cumulativeP \gets cumulativeP + userToItemGraph.WeightOf(e)$
            \If{$p \leq cumulativeP$}
                \State $lastNode \gets currentNode$
                \State $currentNode \gets userToItemGraph.GetSongForEdge(e)$
                \State break
            \EndIf
    \EndFor
\EndFunction
\item[]
\Function{WalkToUser}{$currentNode$, $lastNode$, $userToUserGraph$, $userToItemGraph$}
    \Require $currentNode$ is an Item and $lastNode$ is User with neighbor $currentNode$
    \State $lastNodeNeighbors \gets userToUserGraph.neighborsOf(lastNode)$
    \State $lastNodeNeighborsWithEdgeToCurrentNode \gets filterNodesWithMissingEdge(lastNodeNeighbors, currentNode)$
    \State $songEdgesWeight \gets 0$
    \ForAll{edge $e \in$ \{edges of $currentNode$\}}
            \If{$userToItemGraph.GetUserForEdge(e) \in lastNodeNeighborsWithEdgeToCurrentNode$}
                \State $songEdgesWeight \gets songEdgesWeight + userToItemGraph.WeightOf(e)$
            \EndIf
    \EndFor
    \State $p \gets userEdgesWeight * RandomDoubleBetween(0, 1)$
    \State $cumulativeP \gets 0$
    \ForAll{edge $e \in$ \{edges of $currentNode$\}}
            \If{$userToItemGraph.GetUserForEdge(e) \in lastNodeNeighborsWithEdgeToCurrentNode$}
                \State $cumulativeP \gets cumulativeP + userToItemGraph.WeightOf(e)$
                \If{$p \leq cumulativeP$}
                    \State $lastNode \gets currentNode$
                    \State $currentNode \gets userToItemGraph.GetUserForEdge(e)$
                    \State break
                \EndIf
            \EndIf
    \EndFor
\EndFunction
\end{algorithmic}
\caption{Simplified Web2Recommend SALSA Random Walk}
\label{alg:salsa}
\end{figure*}

\subsection{Timestamp Biased Edge Gossiping} \label{gossiping}

Since Web3Recommend relies on each user storing a local copy of a $TrustNetwork$ each, a mechanism for synchronizing the $TrustNetwork$s across different users is required. Gossiping \cite{jelasity2007gossip} has proved to be a successful mechanism for supporting dynamic and complex information exchange among distributed peers. Gossiping-based mechanisms are great for building and maintaining the network topology and supporting a pervasive diffusion of the information injected into the network. \cite{baraglia2013peer} Gossiping takes inspiration from the human social behavior of spreading information through peers who are in direct contact.

Our gossiping mechanism relies on randomly gossiping user-to-user edges and user-to-item edges to other users in the network. In addition to gossiping, we leverage Semantic Overlay Networks (SON) \cite{crespo2004semantic}, which guarantee that the users we gossip to share similar interests.

GraphJet only generates recommendations based on recent interactions between users and tweets. Thus, only the $n$ latest interactions in the network between users and tweets are stored. Doing this has two advantages:
\begin{enumerate}
\item It allows the creation of temporally contextual (real-time) recommendations which have a profound impact in retail, media, entertainment, and other contexts \cite{ma2020temporal}
\item It provides a mechanism for limiting the memory usage on a device by using $n$ as a tunable hyper-parameter, thus allowing the recommendation algorithm to also be feasible in resource-limited devices such as older smartphones
\end{enumerate}

Web3Recommend achieves this by adding a timestamp to all edges and removing older user-to-item edges after the network's total number of user-to-item edges exceeds $n$. 

Note that while user-to-item edges are deleted after a time window, user-to-user edges are persisted and only deleted either: 1) When they receive gossip with a more recently created version of the edge or 2) The user to user edges of the user exceeds a threshold in which case the edge with the smallest weight is deleted. This is reasonable since, in most media platforms, the number of items dramatically exceeds the number of users; hence, the memory cost of this design choice should be low. Further, it allows us to persist long-term trust relationships between users, which are used in \ref{bootrap}.

We also want to ensure that newly created edges have a higher chance of being gossiped. To do this, we construct a mapping of edges to their delta from the oldest edge in the network. Then, the array is softmax which produces weight values that provide the bias by which an edge is gossiped. Hence, a newer edge is much more likely to be gossiped than an older one.
\subsection{Bootstrap} \label{bootrap}
Recommender Systems suffer from the ``cold start" problem, where the systems meet a new user for the first time. Since the system has no history of the user's interactions, it can't establish the user's personal preference. In a real-time recommendation system like ours, this problem is compounded since an inactive user querying for recommendations may not exist in the interaction graph.

\subsubsection{Circle of Trust}
GraphJet solves this problem by starting the random walks from a \textbf{seed set} instead of a single node. The seed set could be provided by the network or be constructed from the user's trusted users called the ``circle of trust".

In a social network like Twitter, this circle of trust can be calculated using the user's social connections/follows. In our system, we calculate the circle of trust using an Incremental Personalized Page Rank \cite{bahmani2010fast} algorithm that ranks nodes in the network in order of their trustability while ensuring that subsequent random walks are incrementally computed instead of recomputing all the random walks every time edges are updated. 

Since, unlike user-to-item edges, user-to-user edges are persisted over time, inactive users can still be served personalized recommendations in this manner. 

We implement the selection inside the seed set dynamically, ensuring that as the user has more user-to-item interactions, random walks are more likely to start from the user rather than the seed set. Hence, it serves as a bootstrap measure to acquaint the user with the network.

\subsubsection{New User}
The previous bootstrap mechanism relied on the user's prior interactions or social trust in the network. For a fresh user, neither of these exists, and hence a different bootstrap mechanism is required to acquaint them with the network. 

For this, we provide a User Collaborative Filtering algorithm based on the work from \cite{zhang2017improved} for finding similar users in the network. Initially, users are provided with ``divisive" content in the network, which can be used to find more information about their taste profile. This helps establish user-to-song edges that can be used to measure similarity between other users. 

The similarity $sim(a,b)$ between two users $(a, b) \in U$ where $U$ is set of all users is calculated as:
\begin{equation}
    sim(a,b) = Nsim(a,b) \times \tau + (1 - \tau) \times Dsim(a,b)
\end{equation}

Where $Nsim(a,b)$ is calculated as:

\begin{equation} \label{nsim}
    Nsim(a,b) = cf(a,b) \times sim(a,b)
\end{equation}

Here, $sim(a,b)$ is the Pearson Correlation Coefficient \cite{cohen2009pearson} of the two user's item ratings calculated as:

\begin{equation}
    sim(a,b) = \frac{\sum_{i \in I_{ab}}r_{a,i}r_{b,i}}{\sqrt{\sum_{i \in I_{ab}} r_{a,i}^2}\sqrt{\sum_{i \in I_{ab}} r_{b,i}^2}}
\end{equation}

Where $I_{a,b}$ is the set of common rated items between users $a$ and $b$. $r_{a,i} = R_{a,i} - avg_a$, $R_{a,i}$ is the rating given by user $a$ to item $i$ and $avg_a$ the average of all ratings by $a$. 

$cf(a,b)$ is the \textit{common preference degree} between user $a$ and $b$:

\begin{equation}
    cf(a,b) = \frac{\lvert I_a \cap I_b \rvert}{\max_{x \in U} \lvert I_a \cap I_x \rvert}
\end{equation}

$\lvert I_a \cap I_b \rvert$ is the count of common rated items between users $a$ and $b$.

Hence, $Nsim$ \ref{nsim} uses the traditional measure of similarity between users modified by the degree of common preference between the two users. 

$Dsim$ is a measure of similarity calculated using the rating difference of users on their common items (i.e., items that both users have an existing, established affinity for). It helps calculate a more fine-tuned similarity metric than coarsely comparing items both users have interacted with. The exact formula has been omitted for space limitations. The code implementation can be found \href{https://github.com/rmadhwal/trustchain-superapp/blob/TrustedRecommendationsWithExperiments/musicdao/src/main/java/nl/tudelft/trustchain/musicdao/core/recommender/collaborativefiltering/UserBasedTrustedCollaborativeFiltering.kt}{here}. 

Hence, $sim(a,b)$ is used to compare a similarity metric to other users in every user and therefore serves as a bootstrap measure for new users.

Note that this strategy requires that the users being introduced to the new user are non-Sybil. If the strategy is run naively against all users, it is prone to bootstrap threats as mentioned in \ref{applications}.

\subsection{Compact Serialization}
Since Android devices often have limited resources, it's essential to be able to store networks compactly. Once a user disconnects from the network, the \textit{TrustNetwork} is serialized and stored on the device.

Our design is inspired by the \href{https://jgrapht.org/javadoc-1.3.1/org/jgrapht/io/DIMACSFormat.html}{DIMACSFormat} in JGraphT, which is based on the format used in the 2nd DIMACS challenge \cite{johnson1996cliques}.

The details of the serialization have been omitted due to size limitations, but the implementation is self-documenting, rigorously tested, and can be found \href{{https://github.com/rmadhwal/trustchain-superapp/tree/TrustedRecommendations/musicdao/src/main/java/nl/tudelft/trustchain/musicdao/core/recommender/graph/customSerialization}}{here}.

\section{Using MeritRank Decays to generate recommendations} \label{score}
As covered in \ref{meritrank}, MeritRank provides two main mechanisms for Sybil-resistance, Alpha and Beta decays. Each decay is tunable and used to limit the gains of a Sybil Attack.

\subsection{Alpha Decay}
The alpha decay works precisely like the \textit{reset probability} described in \ref{bgt}. The random walk in the user interaction graph stops with probability $\alpha$. Setting an alpha decay reduces the effectiveness of Sybil attacks that rely on long walks that get stuck inside a segment of Sybil users, as described in \cite{meritRank}.

\subsection{Beta Decay}
Beta Decay punishes Sybils for being isolated from the rest of the network. Our system achieves this by measuring the diversity of users voting for a song. 
Assuming the system's beta decay is set to $\beta$, the item beta decay $b[i]$ is calculated for each item $i \in I$ with the following formula:

\begin{equation} \label{beta}
    b[i] = \begin{cases}
        1 - \beta & \text{if } \exists u \in U : div(u,i) > \tau\\
        1 & \text{else }
    \end{cases}
\end{equation}

Where $U$ is the set of all users, $\tau$ is the beta decay threshold and $div(u, i)$ is defined as:
\begin{equation}
    div(u,i) = \frac{\sum_{r \in R(i)} \begin{cases}
        1 & \text{if } \exists u \in U : (u \in r) \cap (r[u] < r[i])) \\
        0 & \text{else }
    \end{cases} } {	\lvert R(i) \rvert}
\end{equation}

Where $R(i)$ is the set of all random walks that contain item $i$, note that SALSA random walks can include both users and items. $r[x \in {U \cup I}]$ is the index of a user or item in a random walk, hence, the step in the random walk when it was walked to. So, each vote item $i$ received in a random walk measures what percentage of times user $u$ led up to the vote. 

Therefore, in \ref{beta} we measure the diversity in recommenders leading up to a vote and compare the ``Sybilness" of item $i$ to the threshold $\tau$ and if it's deemed Sybil, it is assigned a beta decay of $1 - \beta$. 

The rankings of the items in the system are performed using a modified aggregation of random walks. A personalized ranking score $s[i]$ is calculated for each item $i \in I$ using the following formula:
\begin{equation}
    s[i] =  \frac{\lvert R(i) \rvert}{\sum_{x \in I}\lvert R(x) \rvert} b[i]  
\end{equation}

The highest-ranked items are then presented to the user as recommendations.

In our implementation, we provide two methods of calculating beta decays. First, a relatively faster option, which gains its speed by iterating through each random walk linearly and pre-computing the number of times a user is involved in a specific vote for each item. This can be performed in linear runtime relative to the size of random walks. But the space cost of the algorithm makes it prohibitively expensive on devices with limited memory. Hence, we also provide an on-the-fly implementation that is more time-consuming but has a tiny memory footprint.

\section{Experiment Setup} \label{experimentsetup}
\subsection{Dataset and Test Network Generation}
To demonstrate Sybil resistance, it was important to create a network of non-Sybil users on which we could mount Sybil attacks. We used the dataset from the taste profile subset of the Million Song Dataset \cite{bertin2011million,10.1145/2187980.2188222} to generate this non-Sybil network. The original dataset was sourced from The Echo Nest, an online resource that provides music applications on the web, smartphones, etc. Therefore, using this data from real music platforms, we could generate a network that emulates the behavior of non-Sybil users in Web3Recommend.
The taste profile subset provides us with:
\begin{enumerate}
    \item Users
    \item Songs
    \item User-Song-Play Count Triplet
\end{enumerate}

For the test network generation, user-song edges were created using the user-song-play count triplet to estimate user-song affinity as described in \ref{affinitytrust}. User-User edges were created using the same user-based collaborative filtering method we used to bootstrap trust relationships for new users. This is reasonable since all the users in this network were assumed to be non-Sybil; hence, the trust between them is simply a function of their similarity. 

\begin{figure*}[t]
    \centering
    \captionsetup{format=myformat}
    \includegraphics[width=0.7\textwidth]{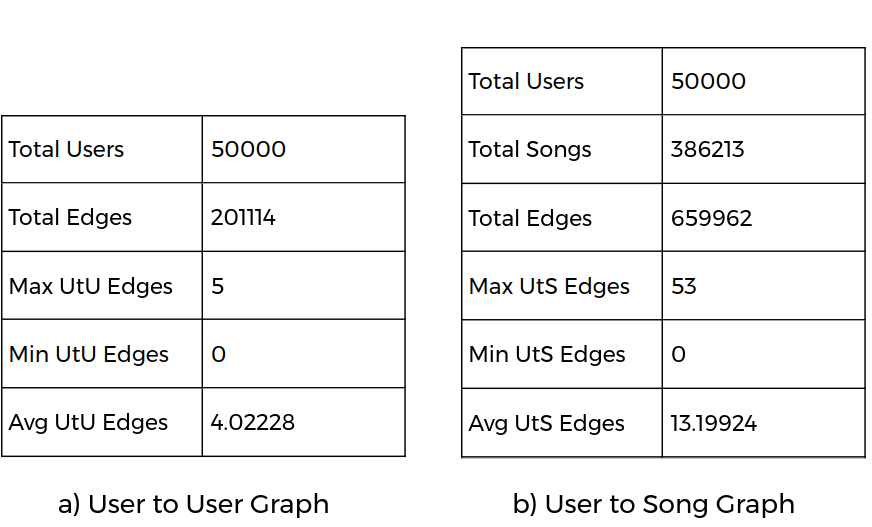}
    \caption{Test Network Stats (Note that UtU stands for User to User and UtS stands for User to Song)}
    \label{fig:graph_stats}
\end{figure*}

Figure \ref{fig:graph_stats} shows some statistics of the test network.

\subsection{Metrics used for experiments}
\subsubsection{Top x Ranking}
The goal of Web3Recommend is to rank items available to a user in order of increasing relevance and trust to them. However, in most consumer-driven recommendation systems, only the top few recommendations are important since those are the only ones the user would usually look at to choose the next item to consume. Therefore, in most of our experiments, we measure the influence on the system's top 100/1000/10000 ranked items.

Specifically, for the experiments which measure Sybil resistance we were interested in measuring how many Sybil songs ended up in these rankings. For example, a Sybil item being the highest-ranked recommendation to a user is much worse than two Sybil items ranked much lower. Therefore, we define a metric \textit{Sybil influence of top $x$ ranking} or $SITR(x)$, which measures the influence of Sybils in the top $x$ category of ranking using the follow formula:

\begin{equation}
    SITR(x) = \sum^{i=x}_{i = 0} \begin{cases}
        x-i & \text{if } DRankedI(i) \in \mathbb{S}\\
        0 & \text{else }
    \end{cases}
\end{equation}

Where, $DRankedI$ is the array of all items sorted in descending order of their ranking and $\mathbb{S}$ is the set of all Sybil items. Hence, for example, if there is only one Sybil item in the network but it is the highest ranked item, the $SITR(100)$ will be 100.

\subsubsection{Ranked Biased Overlap}
In order to measure the effect of decay parameters on the ranking of the items, we use Ranked Biased Overlap \cite{webber2010similarity} between two sets of ranking to compare how similar they are. Ranked Biased Overlap is specifically constructed to be a similarity measure between incomplete ranking lists and hence fits our use case very well. RBO outputs a value between 0.0 and 1.0, where 1.0 represents two identical sets while 0.0 represents completely different sets.

\section{Experiments} \label{experiments}
\subsection{Leave one out cross validation}
In this set of experiments, we follow the below process:
\begin{enumerate}
    \item 100 users from the User to Song graph are sampled randomly
    \item For each user, we remove an existing user to song edge
    \item We run the personalized SALSA algorithm with alpha decay value set to $\alpha$
    \item We measure: 1) The ranking score of the song whose edge was removed 2) In how many percent of trials the missing song shows up in the Top 100/1000/10000 rankings
\end{enumerate}

While the goal of our paper is focused on limiting Sybil influence, this experiment helps demonstrate that the foundational recommendation algorithm can provide reasonable personalized recommendations. The result of our algorithm is also compared to the result of a simple vanilla personalized SALSA implementation to show that our algorithm performs just as well, if not better in terms of recommendations for our dataset.

\subsection{Effect on the ranking of increasing decays}
In this set of experiments, we follow the below process:
\begin{enumerate}
    \item 100 users from the User to Song graph are sampled randomly
    \item For each user, we run the personalized SALSA algorithm for a specific alpha decay value $\alpha \in {0.1,1.0}$/beta decay value $\beta \in {0.1,1.0}$
    \item For each ranking with increasing $\alpha$ or $\beta$ values, we calculate the Ranked Biased Offset compared to the original ranking
\end{enumerate}

This experiment helps to measure the influence on ranking with increasing decay values, showing us how growing decays in the network and hence, increasing trust by reducing the influence of Sybils impacts the relevance of recommendations. This could also be interpreted as the ``false positive" rate of Sybil detection.

\subsection{Single Sybil attack}
As mentioned earlier, MeritRank provides Sybil resistance by limiting Sybil influence through:
\begin{equation}
\lim_{|S|\to\infty} \frac{w^+(\sigma_s)}{w^-(\sigma_s)} \leq c \\
\end{equation}

In this experiment, we prove this property in our network by constructing multiple Sybil attacks where an adversary with an existing trust edge to non-Sybil users starts mounting a Sybil attack by creating multiple Sybil identities. Thus, if we can prove that after a certain threshold of identities, the benefit to the adversary of creating more Sybils is negligible, we can show that our system offers Sybil resistance.

Therefore, in this experiment, we follow the below process:
\begin{enumerate}
    \item In our test network, we randomly sample a user
    \item A trusted neighbor of the user is converted to a \textbf{traitor} who mounts either a: 1) linear Sybil attack or 2) parallel Sybil attack as defined in \ref{meritrank}
    \item The traitor mounts attacks with an increasing number of Sybils and for each attack we measure the gain in total ranking score in our personalized SALSA algorithm of the Sybil items created by the adversary
\end{enumerate}

\subsection{Giga Sybil attack}

In a real network, multiple users could potentially mount a Sybil attack. Hence, it is worth evaluating the performance of our network as an increasing percentage of all users in the network become malicious users. For the first half of the experiments, we set 50\% of all the nodes to Sybil nodes and measured the influence gained for Sybil songs over sampled users with varying decay parameters like in the previous experiments. This allows us to measure the gain for Sybils from a ``giga Sybil" attack with different decay parameters.

However, after the \% Sybil users in the network pass a certain threshold, our system's ranking mechanism can't work effectively since the only items that random walks could potentially discover would be Sybil items and hence, due to the influence of beta decays, the rankings produced by our system will be effectively random. Our experiments found this threshold to be around $0.6$ for the chosen dataset.

\section{Experiment Results} \label{er}
\subsection{Leave one out cross validation}
\begin{figure}[H]
    \centering
    \captionsetup{format=myformat}
    \includegraphics[width=0.5\textwidth]{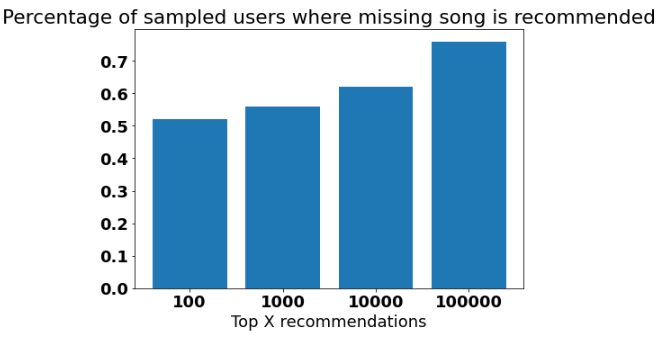}
    \caption{Leave out one experiment with $\alpha$ = 0.1 and $\beta = 0.0$}
    \label{1e}
\end{figure}

As shown in \ref{1e} with alpha decay set to 0.1, in more than half of the sampled users, the missing song is within the top 100 recommendations (the dataset consists of 386213 songs, so this is top 0.0003\% of the results) and about 80\% of the times it is in the top 10000 (top 33\% recommendations).

\begin{figure}[H]
    \captionsetup{format=myformat}
    \includegraphics[width=0.5\textwidth]{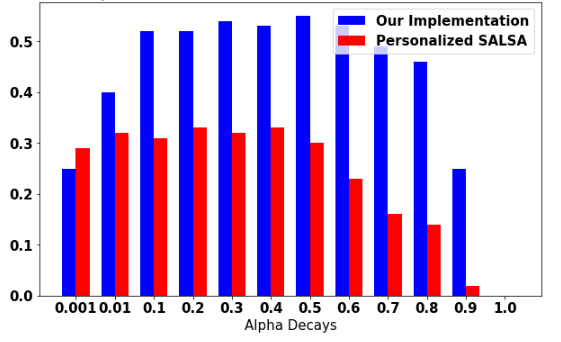}
    \caption{\% of missing songs in top 100 recommendations in both strategies with varying levels of alpha decay}
    \label{2e}
\end{figure}

\begin{figure}[H]
    \centering
    \captionsetup{format=myformat}
    \includegraphics[width=0.5\textwidth]{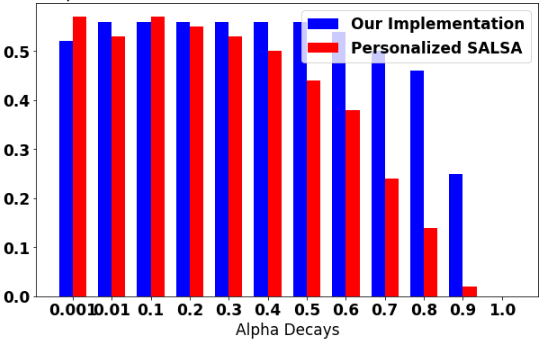}
    \caption{\% of missing songs in top 1000 recommendations in both strategies with varying levels of alpha decay}
    \label{3e}
\end{figure}

\begin{figure}[H]
    \centering
    \captionsetup{format=myformat}
    \includegraphics[width=0.5\textwidth]{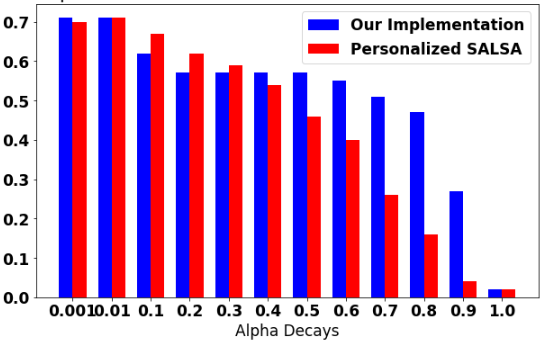}
    \caption{\% of missing songs in top 10000 recommendations in both strategies with varying levels of alpha decay}
    \label{4e}
\end{figure}

Next, we compare our system's performance to a vanilla personalized SALSA recommendation system that doesn't include any enhancements mentioned in \ref{systemdesign}. Figures \ref{2e}, \ref{3e}, and \ref{4e} compare the performance of both systems in measuring \% of missing songs in top X recommendations. While in the top 1000 and 10000 recommendations, our system performs similarly to the vanilla algorithm, in the top 100 recommendations, it performs much better. We believe that the top 100 recommendations metric is also the most significant since users often only look at the highest recommendations.

\subsection{Effect on the ranking of increasing decays}

\begin{figure}[H]
    \centering
    \captionsetup{format=myformat}
    \includegraphics[width=0.5\textwidth]{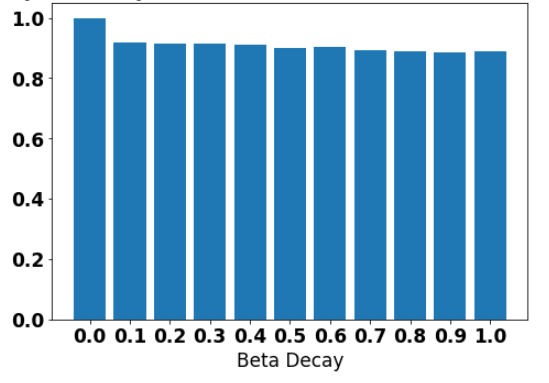}
    \caption{RBO similarity of non-Sybil recommendations with increasing Beta Decay}
    \label{5e}
\end{figure}

\begin{figure}[H]
    \centering
    \captionsetup{format=myformat}
    \includegraphics[width=0.5\textwidth]{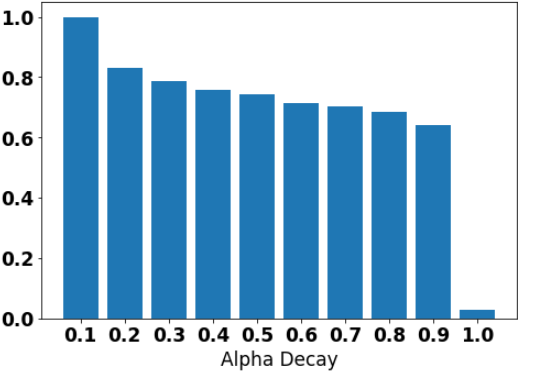}
    \caption{RBO similarity of non-Sybil recommendations with increasing Alpha Decay}
    \label{6e}
\end{figure}

As shown in figures \ref{5e} and \ref{6e}, the false positive impact increases slightly with increasing Alpha Decay while increasing Beta Decay seems to have an almost negligible impact. Thus, a real implementation would be better suited to use a low Alpha Decay and a high Beta Decay. In the MusicDAO implementation, we set Alpha Decay to $0.1$ and Beta Decay to $0.8$.

\subsection{Single Sybil attack}

\begin{figure}[H]
    \centering
    \captionsetup{format=myformat}
    \includegraphics[width=0.5\textwidth]{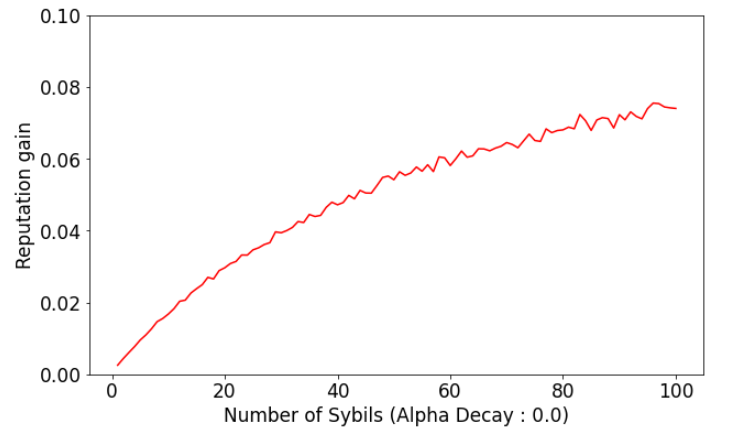}
    \caption{Ranking score gained for a malicious node with increasingly larger linear Sybil attacks (Alpha Decay: 0.0)}
    \label{7e}
\end{figure}

\begin{figure}[H]
    \centering
    \captionsetup{format=myformat}
    \includegraphics[width=0.5\textwidth]{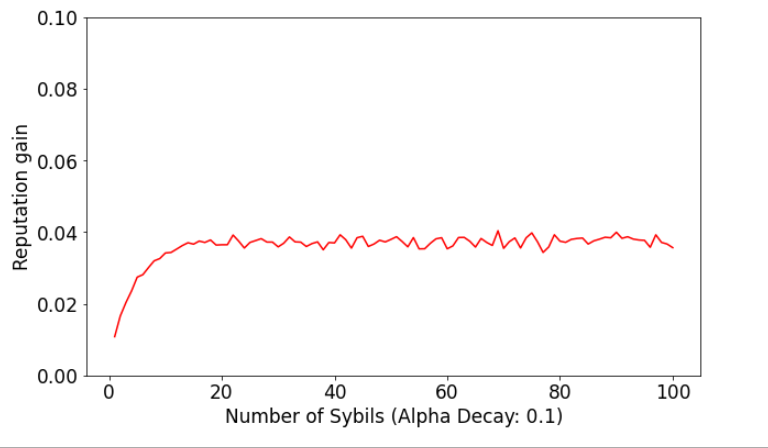}
    \caption{Ranking score gained for a malicious node with increasingly larger linear Sybil attacks (Alpha Decay: 0.1)}
    \label{8e}
\end{figure}

\begin{figure}[H]
    \centering
    \captionsetup{format=myformat}
    \includegraphics[width=0.5\textwidth]{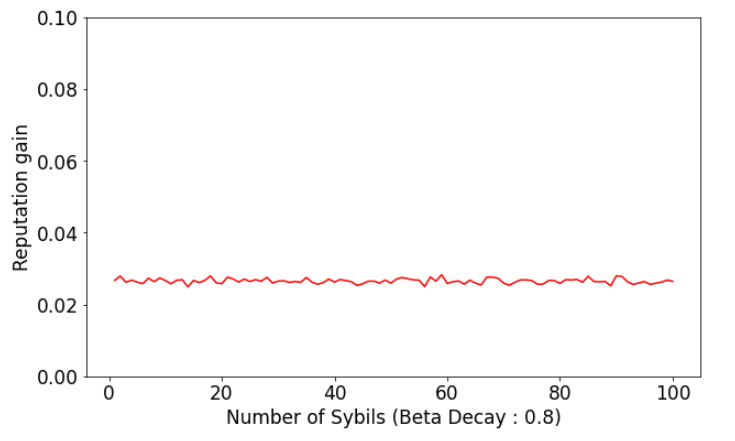}
    \caption{Ranking score gained for a malicious node with increasingly larger parallel Sybil attacks (Beta Decay: 0.8)}
    \label{9e}
\end{figure}

First, in \ref{7e}, we show how an adversary can gain a theoretically infinite amount of ranking scores through linear Sybil attacks without Alpha Decay. Then in \ref{8e}, we rerun the experiment with Alpha Decay set to $0.1$. After a certain number of Sybils, the reputation gain for the adversary remains the same, hence creating more Sybils doesn't provide any additional benefit to the attacker. In \ref{9e}, we demonstrate the same effect with Beta Decay set to $0.8$ against parallel attacks.

\subsection{Giga Sybil attack}

\begin{figure}[H]
    \centering
    \captionsetup{format=myformat}
    \includegraphics[width=0.5\textwidth]{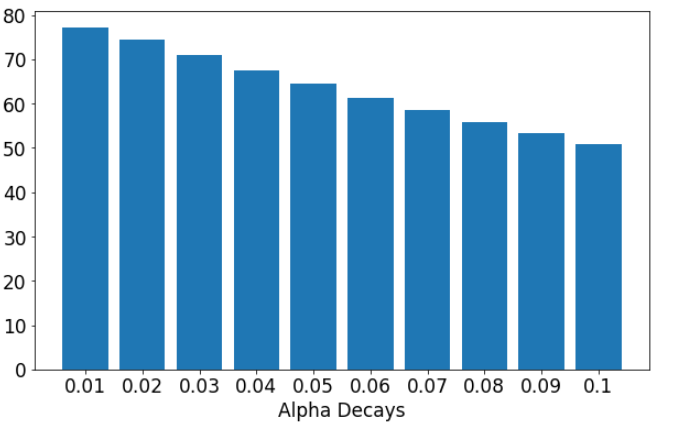}
    \caption{Cumulative Score for Sybil songs in Giga Sybil Attack with increasing Alpha Decay}
    \label{10e}
\end{figure}

\begin{figure}[H]
    \centering
    \captionsetup{format=myformat}
    \includegraphics[width=0.5\textwidth]{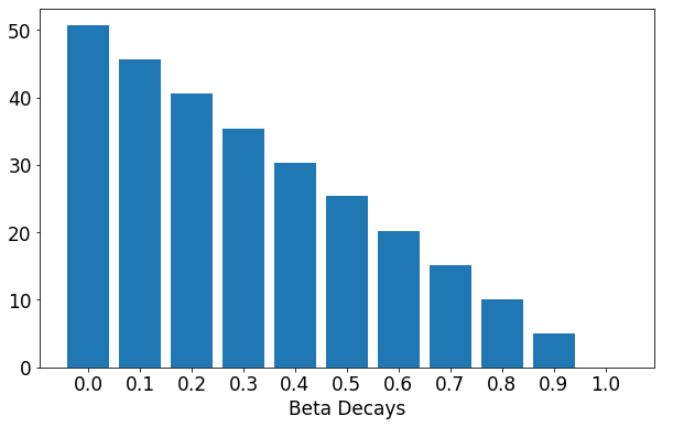}
    \caption{Cumulative Score for Sybil songs in Giga Sybil Attack with increasing Beta Decay}
    \label{11e}
\end{figure}

Figures \ref{10e} and \ref{11e} show the change in combined ranking score for Sybil songs in our Giga Sybil Attack with varying Alpha and Beta decays respectively. Both mechanisms are able to limit Sybil gain as expected.

\begin{figure}[H]
    \centering
    \captionsetup{format=myformat}
    \includegraphics[width=0.5\textwidth]{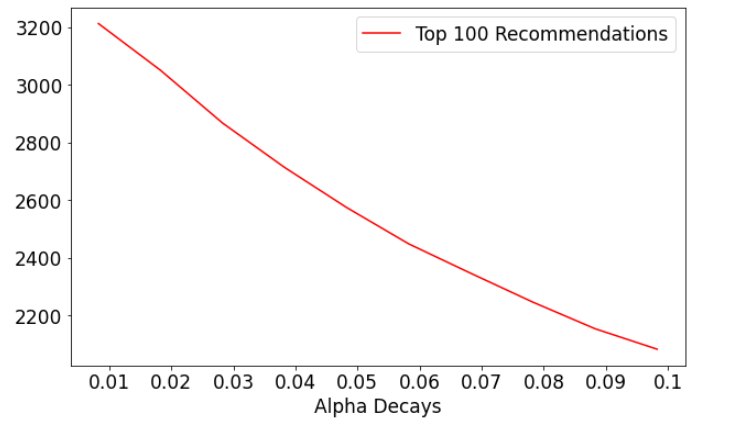}
    \caption{SITR(100) for Sybil songs in Giga Sybil Attack with increasing Alpha Decay}
    \label{12e}
\end{figure}

\begin{figure}[H]
    \centering
    \captionsetup{format=myformat}
    \includegraphics[width=0.5\textwidth]{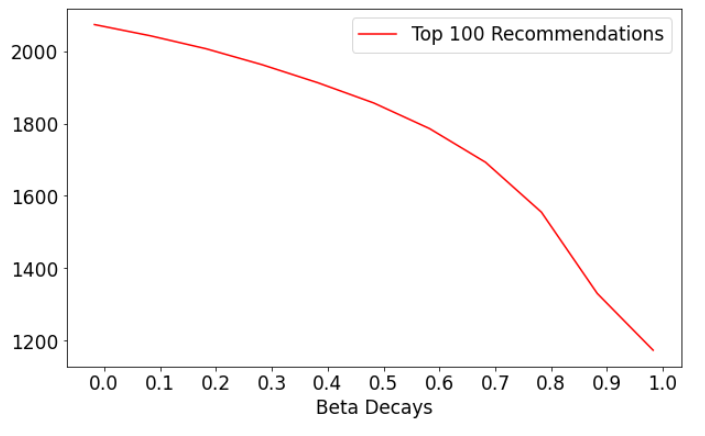}
    \caption{SITR(100) for Sybil songs in Giga Sybil Attack with increasing Beta Decay (Alpha Decay:0.1)}
    \label{13e}
\end{figure}

Figures \ref{12e} and \ref{13e} show the effect of increasing the decay values on the SITR(100). Note that even though with beta decays, we can completely eliminate Sybil attack edges, in a giga Sybil attack, the majority of the items detected are Sybil items, and hence even though their score is $0$, since non-Sybil songs were starved of random walks, they have a much lower score too. However, as shown in the graphs, adding the decay values can noticeably reduce the giga Sybil attack's gain even if it doesn't eliminate it.

\section{Testing} \label{test}
All the mentioned components of Web3Recommend's Kotlin implementation are tested using JUnit.

\section{Real World Deployment} \label{rwd}
Outside experiments, as a proof of concept, Web3Recommend was also integrated into MusicDAO, a peer-to-peer music-sharing application that aims to rival Spotify. Figure \ref{fig:musicdao} shows recommendations being generated in the application in a network of 3 connected users. The code for MusicDAO is open source and it is freely available to be downloaded through the Android store.

\section{Further Work} \label{furtherwork}
Future research directions could explore further improvements to the recommendation algorithm, such as enhancing the scalability of the graph-based algorithms.

We acknowledge that in our paper the definiton of ``trust" is simply limited to Sybil-resistance, while in the real world, untrustworthy content also includes content that is malicious in other ways such as through being divisive, containing unfactual content (fake news), being gory, part of a scam/ponzi scheme, etc. Investigating the integration of additional reputation mechanisms could increase trust beyond simple Sybil-resistance 

Further real-world deployments could be used to evaluate the system's performance and usability in more diverse settings than only in the context of music sharing. 

Finally, in \ref{smaa} we list a lot of assumptions and limitations stated in the system and how future implementations can address these.

\section{Conclusion} \label{conc}
In this paper, we presented Web3Recommend, a decentralized Social Recommender System designed to generate trustworthy and relevant recommendations in Web3 platforms on Android. We addressed the challenges posed by decentralized networks, such as the lack of a global perspective and susceptibility to Sybil Attacks.

By integrating the MeritRank decentralized reputation scheme into our graph-based recommendation design, we achieved Sybil-resistance in the generated recommendations. Our experiments included evaluations against multiple adversarial strategies. The results demonstrated the trust-relevance balance of our recommendations, showcasing the system's ability to generate personalized, real-time recommendations.

In summary, Web3Recommend represents a significant advancement in Social Recommender Systems. Combining decentralized network principles, the MeritRank reputation scheme, and efficient graph-based algorithms, we have created a Sybil-resistant recommendation system capable of generating real-time recommendations in Web3 platforms. Our work contributes to the research community and offers practical benefits to users by enhancing their Web3 platform experience.

With the increasing adoption of Web3 technologies, the development of trustworthy and relevant recommender systems will continue to play a crucial role in enabling users to discover valuable content and build meaningful connections within Web3 platforms.

\bibliographystyle{IEEEtran}
\bibliography{references}

\begin{thebibliography}{10}
\providecommand{\url}[1]{#1}
\csname url@samestyle\endcsname
\providecommand{\newblock}{\relax}
\providecommand{\bibinfo}[2]{#2}
\providecommand{\BIBentrySTDinterwordspacing}{\spaceskip=0pt\relax}
\providecommand{\BIBentryALTinterwordstretchfactor}{4}
\providecommand{\BIBentryALTinterwordspacing}{\spaceskip=\fontdimen2\font plus
\BIBentryALTinterwordstretchfactor\fontdimen3\font minus
  \fontdimen4\font\relax}
\providecommand{\BIBforeignlanguage}[2]{{%
\expandafter\ifx\csname l@#1\endcsname\relax
\typeout{** WARNING: IEEEtran.bst: No hyphenation pattern has been}%
\typeout{** loaded for the language `#1'. Using the pattern for}%
\typeout{** the default language instead.}%
\else
\language=\csname l@#1\endcsname
\fi
#2}}
\providecommand{\BIBdecl}{\relax}
\BIBdecl

\bibitem{lu2014toward}
R.~Lu, H.~Zhu, X.~Liu, J.~K. Liu, and J.~Shao, ``Toward efficient and
  privacy-preserving computing in big data era,'' \emph{IEEE Network}, vol.~28,
  no.~4, pp. 46--50, 2014.

\bibitem{isinkaye2015recommendation}
F.~O. Isinkaye, Y.~O. Folajimi, and B.~A. Ojokoh, ``Recommendation systems:
  Principles, methods and evaluation,'' \emph{Egyptian informatics journal},
  vol.~16, no.~3, pp. 261--273, 2015.

\bibitem{shenk1999data}
D.~Shenk, ``Data smog: Surviving the information glut,'' 1999.

\bibitem{levitin2014organized}
D.~J. Levitin, \emph{The organized mind: Thinking straight in the age of
  information overload}.\hskip 1em plus 0.5em minus 0.4em\relax Penguin, 2014.

\bibitem{bawden2020information}
D.~Bawden and L.~Robinson, ``Information overload: An overview,'' 2020.

\bibitem{deathByIO}
P.~Hemp, ``Death by information overload,'' \emph{Harvard business review},
  vol.~87, pp. 82--9, 121, 10 2009.

\bibitem{tiktokstats}
\BIBentryALTinterwordspacing
Admin, ``Tiktok statistics - everything you need to know [mar 2023 update],''
  Mar 2023. [Online]. Available:
  \url{https://wallaroomedia.com/blog/social-media/tiktok-statistics/}
\BIBentrySTDinterwordspacing

\bibitem{gupta2013wtf}
P.~Gupta, A.~Goel, J.~Lin, A.~Sharma, D.~Wang, and R.~Zadeh, ``Wtf: The who to
  follow service at twitter,'' in \emph{Proceedings of the 22nd international
  conference on World Wide Web}, 2013, pp. 505--514.

\bibitem{das2017survey}
D.~Das, L.~Sahoo, and S.~Datta, ``A survey on recommendation system,''
  \emph{International Journal of Computer Applications}, vol. 160, no.~7, 2017.

\bibitem{guy2011social}
I.~Guy and D.~Carmel, ``Social recommender systems,'' in \emph{Proceedings of
  the 20th international conference companion on World wide web}, 2011, pp.
  283--284.

\bibitem{sharma2016graphjet}
A.~Sharma, J.~Jiang, P.~Bommannavar, B.~Larson, and J.~Lin, ``Graphjet:
  Real-time content recommendations at twitter,'' \emph{Proceedings of the VLDB
  Endowment}, vol.~9, no.~13, pp. 1281--1292, 2016.

\bibitem{lempel2001salsa}
R.~Lempel and S.~Moran, ``Salsa: the stochastic approach for link-structure
  analysis,'' \emph{ACM Transactions on Information Systems (TOIS)}, vol.~19,
  no.~2, pp. 131--160, 2001.

\bibitem{chaabane2014closer}
A.~Chaabane, Y.~Ding, R.~Dey, M.~A. Kaafar, and K.~W. Ross, ``A closer look at
  third-party osn applications: are they leaking your personal information?''
  in \emph{Passive and Active Measurement: 15th International Conference, PAM
  2014, Los Angeles, CA, USA, March 10-11, 2014, Proceedings 15}.\hskip 1em
  plus 0.5em minus 0.4em\relax Springer, 2014, pp. 235--246.

\bibitem{hassan2017replication}
A.~Hassan, ``Replication and availability in decentralised online social
  networks,'' 2017.

\bibitem{bambacht2022web3}
J.~Bambacht and J.~Pouwelse, ``Web3: A decentralized societal infrastructure
  for identity, trust, money, and data,'' \emph{arXiv preprint
  arXiv:2203.00398}, 2022.

\bibitem{nasrulin2022meritrank}
B.~Nasrulin, G.~Ishmaev, and J.~Pouwelse, ``Meritrank: Sybil tolerant
  reputation for merit-based tokenomics,'' in \emph{2022 4th Conference on
  Blockchain Research \& Applications for Innovative Networks and Services
  (BRAINS)}.\hskip 1em plus 0.5em minus 0.4em\relax IEEE, 2022, pp. 95--102.

\bibitem{wissel2021fairness}
T.~Wissel, ``Fairness and freedom for artists: Towards a robot economy for the
  music industry,'' 2021.

\bibitem{gray1986approach}
J.~N. Gray, ``An approach to decentralized computer systems,'' \emph{IEEE
  Transactions on Software Engineering}, no.~6, pp. 684--692, 1986.

\bibitem{douceur2002sybil}
J.~R. Douceur, ``The sybil attack,'' in \emph{Peer-to-Peer Systems: First
  InternationalWorkshop, IPTPS 2002 Cambridge, MA, USA, March 7--8, 2002
  Revised Papers 1}.\hskip 1em plus 0.5em minus 0.4em\relax Springer, 2002, pp.
  251--260.

\bibitem{apte2019frauds}
M.~Apte, G.~K. Palshikar, and S.~Baskaran, ``Frauds in online social networks:
  A review,'' \emph{Social networks and surveillance for society}, pp. 1--18,
  2019.

\bibitem{Contributor_2007}
\BIBentryALTinterwordspacing
Contributor, ``Stat gaming services come to youtube,'' Aug 2007. [Online].
  Available:
  \url{https://techcrunch.com/2007/08/23/myspace-style-profile-gaming-comes-to-youtube/}
\BIBentrySTDinterwordspacing

\bibitem{yu2009dsybil}
H.~Yu, C.~Shi, M.~Kaminsky, P.~B. Gibbons, and F.~Xiao, ``Dsybil: Optimal
  sybil-resistance for recommendation systems,'' in \emph{2009 30th IEEE
  Symposium on Security and Privacy}.\hskip 1em plus 0.5em minus 0.4em\relax
  IEEE, 2009, pp. 283--298.

\bibitem{avrachenkov2007monte}
K.~Avrachenkov, N.~Litvak, D.~Nemirovsky, and N.~Osipova, ``Monte carlo methods
  in pagerank computation: When one iteration is sufficient,'' \emph{SIAM
  Journal on Numerical Analysis}, vol.~45, no.~2, pp. 890--904, 2007.

\bibitem{leskovec2008community}
J.~Leskovec, K.~J. Lang, A.~Dasgupta, and M.~W. Mahoney, ``Community structure
  in large networks: Natural cluster sizes and the absence of large
  well-defined clusters,'' 2008.

\bibitem{rashid2002getting}
A.~M. Rashid, I.~Albert, D.~Cosley, S.~K. Lam, S.~M. McNee, J.~A. Konstan, and
  J.~Riedl, ``Getting to know you: learning new user preferences in recommender
  systems,'' in \emph{Proceedings of the 7th international conference on
  Intelligent user interfaces}, 2002, pp. 127--134.

\bibitem{page1999pagerank}
L.~Page, S.~Brin, R.~Motwani, and T.~Winograd, ``The pagerank citation ranking:
  Bringing order to the web.'' Stanford InfoLab, Tech. Rep., 1999.

\bibitem{10.1145/371920.372096}
\BIBentryALTinterwordspacing
A.~Borodin, G.~O. Roberts, J.~S. Rosenthal, and P.~Tsaparas, ``Finding
  authorities and hubs from link structures on the world wide web,'' in
  \emph{Proceedings of the 10th International Conference on World Wide Web},
  ser. WWW '01.\hskip 1em plus 0.5em minus 0.4em\relax New York, NY, USA:
  Association for Computing Machinery, 2001, p. 415–429. [Online]. Available:
  \url{https://doi-org.tudelft.idm.oclc.org/10.1145/371920.372096}
\BIBentrySTDinterwordspacing

\bibitem{fogaras2005towards}
D.~Fogaras, B.~R{\'a}cz, K.~Csalog{\'a}ny, and T.~Sarl{\'o}s, ``Towards scaling
  fully personalized pagerank: Algorithms, lower bounds, and experiments,''
  \emph{Internet Mathematics}, vol.~2, no.~3, pp. 333--358, 2005.

\bibitem{blum2006random}
A.~Blum, T.~H. Chan, and M.~R. Rwebangira, ``A random-surfer web-graph model,''
  in \emph{2006 Proceedings of the Third Workshop on Analytic Algorithmics and
  Combinatorics (ANALCO)}.\hskip 1em plus 0.5em minus 0.4em\relax SIAM, 2006,
  pp. 238--246.

\bibitem{bahmani2010fast}
B.~Bahmani, A.~Chowdhury, and A.~Goel, ``Fast incremental and personalized
  pagerank,'' \emph{arXiv preprint arXiv:1006.2880}, 2010.

\bibitem{kleinberg1999authoritative}
J.~M. Kleinberg, ``Authoritative sources in a hyperlinked environment,''
  \emph{Journal of the ACM (JACM)}, vol.~46, no.~5, pp. 604--632, 1999.

\bibitem{baran}
P.~Baran, ``On distributed communications networks,'' \emph{IEEE Transactions
  on Communications Systems}, vol.~12, no.~1, pp. 1--9, 1964.

\bibitem{Korpal2022}
\BIBentryALTinterwordspacing
G.~Korpal and D.~Scott, ``{Decentralization and web3 technologies},'' 5 2022.
  [Online]. Available:
  \url{https://www.techrxiv.org/articles/preprint/Decentralization_and_web3_technologies/19727734}
\BIBentrySTDinterwordspacing

\bibitem{tribler}
J.~Pouwelse, P.~Garbacki, J.~Wang, A.~Bakker, J.~Yang, A.~Iosup, D.~Epema,
  M.~Reinders, M.~van Steen, and H.~Sips, ``Tribler: a social‐based
  peer‐to‐peer system,'' \emph{Concurrency and Computation: Practice and
  Experience}, vol.~20, pp. 127 -- 138, 02 2008.

\bibitem{o2007web}
T.~O'reilly, ``What is web 2.0: Design patterns and business models for the
  next generation of software,'' \emph{Communications \& strategies}, no.~1,
  p.~17, 2007.

\bibitem{solid}
\BIBentryALTinterwordspacing
``Home · solid.'' [Online]. Available: \url{https://solidproject.org/}
\BIBentrySTDinterwordspacing

\bibitem{wood}
\BIBentryALTinterwordspacing
G.~Wood. [Online]. Available: \url{http://gavwood.com/dappsweb3.html}
\BIBentrySTDinterwordspacing

\bibitem{eigenTrust}
S.~Kamvar, M.~Schlosser, and H.~Garcia-molina, ``The eigentrust algorithm for
  reputation management in p2p networks,'' \emph{The EigenTrust Algorithm for
  Reputation Management in P2P Networks}, 04 2003.

\bibitem{kollock1999production}
P.~Kollock \emph{et~al.}, ``The production of trust in online markets,''
  \emph{Advances in group processes}, vol.~16, no.~1, pp. 99--123, 1999.

\bibitem{meritRank}
\BIBentryALTinterwordspacing
B.~Nasrulin, G.~Ishmaev, and J.~Pouwelse, ``Meritrank: Sybil tolerant
  reputation for merit-based tokenomics,'' 2022. [Online]. Available:
  \url{https://arxiv.org/abs/2207.09950}
\BIBentrySTDinterwordspacing

\bibitem{borisov2006computational}
N.~Borisov, ``Computational puzzles as sybil defenses,'' in \emph{Sixth IEEE
  International Conference on Peer-to-Peer Computing (P2P'06)}.\hskip 1em plus
  0.5em minus 0.4em\relax IEEE, 2006, pp. 171--176.

\bibitem{damiani2002reputation}
E.~Damiani, D.~C. di~Vimercati, S.~Paraboschi, P.~Samarati, and F.~Violante,
  ``A reputation-based approach for choosing reliable resources in peer-to-peer
  networks,'' in \emph{Proceedings of the 9th ACM conference on Computer and
  communications security}, 2002, pp. 207--216.

\bibitem{rowaihy2007limiting}
H.~Rowaihy, W.~Enck, P.~McDaniel, and T.~La~Porta, ``Limiting sybil attacks in
  structured p2p networks,'' in \emph{IEEE INFOCOM 2007-26th IEEE International
  Conference on Computer Communications}.\hskip 1em plus 0.5em minus
  0.4em\relax IEEE, 2007, pp. 2596--2600.

\bibitem{bazzi2005establishment}
R.~A. Bazzi and G.~Konjevod, ``On the establishment of distinct identities in
  overlay networks,'' in \emph{Proceedings of the twenty-fourth annual ACM
  symposium on Principles of distributed computing}, 2005, pp. 312--320.

\bibitem{resnick2000reputation}
P.~Resnick, K.~Kuwabara, R.~Zeckhauser, and E.~Friedman, ``Reputation
  systems,'' \emph{Communications of the ACM}, vol.~43, no.~12, pp. 45--48,
  2000.

\bibitem{resnick2006value}
P.~Resnick, R.~Zeckhauser, J.~Swanson, and K.~Lockwood, ``The value of
  reputation on ebay: A controlled experiment,'' \emph{Experimental economics},
  vol.~9, pp. 79--101, 2006.

\bibitem{yu2006sybilguard}
H.~Yu, M.~Kaminsky, P.~B. Gibbons, and A.~Flaxman, ``Sybilguard: defending
  against sybil attacks via social networks,'' in \emph{Proceedings of the 2006
  conference on Applications, technologies, architectures, and protocols for
  computer communications}, 2006, pp. 267--278.

\bibitem{yu2008sybillimit}
H.~Yu, P.~B. Gibbons, M.~Kaminsky, and F.~Xiao, ``Sybillimit: A near-optimal
  social network defense against sybil attacks,'' in \emph{2008 IEEE Symposium
  on Security and Privacy (sp 2008)}.\hskip 1em plus 0.5em minus 0.4em\relax
  IEEE, 2008, pp. 3--17.

\bibitem{mislove2008ostra}
A.~Mislove, A.~Post, P.~Druschel, and P.~K. Gummadi, ``Ostra: Leveraging trust
  to thwart unwanted communication.'' in \emph{Nsdi}, vol.~8, 2008, pp. 15--30.

\bibitem{tran2009sybil}
D.~N. Tran, B.~Min, J.~Li, and L.~Subramanian, ``Sybil-resilient online content
  voting.'' in \emph{NSDI}, vol.~9, no.~1, 2009, pp. 15--28.

\bibitem{auer2002nonstochastic}
P.~Auer, N.~Cesa-Bianchi, Y.~Freund, and R.~E. Schapire, ``The nonstochastic
  multiarmed bandit problem,'' \emph{SIAM journal on computing}, vol.~32,
  no.~1, pp. 48--77, 2002.

\bibitem{li2010contextual}
L.~Li, W.~Chu, J.~Langford, and R.~E. Schapire, ``A contextual-bandit approach
  to personalized news article recommendation,'' in \emph{Proceedings of the
  19th international conference on World wide web}, 2010, pp. 661--670.

\bibitem{guo2015learning}
L.~Guo, J.~Ma, Z.~Chen, and H.~Zhong, ``Learning to recommend with social
  contextual information from implicit feedback,'' \emph{Soft Computing},
  vol.~19, pp. 1351--1362, 2015.

\bibitem{bobadilla2013recommender}
J.~Bobadilla, F.~Ortega, A.~Hernando, and A.~Guti{\'e}rrez, ``Recommender
  systems survey,'' \emph{Knowledge-based systems}, vol.~46, pp. 109--132,
  2013.

\bibitem{kurdi2015honestpeer}
H.~A. Kurdi, ``Honestpeer: An enhanced eigentrust algorithm for reputation
  management in p2p systems,'' \emph{Journal of King Saud University-Computer
  and Information Sciences}, vol.~27, no.~3, pp. 315--322, 2015.

\bibitem{ripeanu2001peer}
M.~Ripeanu, ``Peer-to-peer architecture case study: Gnutella network,'' in
  \emph{Proceedings first international conference on peer-to-peer
  computing}.\hskip 1em plus 0.5em minus 0.4em\relax IEEE, 2001, pp. 99--100.

\bibitem{michail2020jgrapht}
D.~Michail, J.~Kinable, B.~Naveh, and J.~V. Sichi, ``Jgrapht—a java library
  for graph data structures and algorithms,'' \emph{ACM Transactions on
  Mathematical Software (TOMS)}, vol.~46, no.~2, pp. 1--29, 2020.

\bibitem{jelasity2007gossip}
M.~Jelasity, S.~Voulgaris, R.~Guerraoui, A.-M. Kermarrec, and M.~Van~Steen,
  ``Gossip-based peer sampling,'' \emph{ACM Transactions on Computer Systems
  (TOCS)}, vol.~25, no.~3, pp. 8--es, 2007.

\bibitem{baraglia2013peer}
R.~Baraglia, P.~Dazzi, M.~Mordacchini, and L.~Ricci, ``A peer-to-peer
  recommender system for self-emerging user communities based on gossip
  overlays,'' \emph{Journal of Computer and System Sciences}, vol.~79, no.~2,
  pp. 291--308, 2013.

\bibitem{crespo2004semantic}
A.~Crespo and H.~Garcia-Molina, ``Semantic overlay networks for p2p systems,''
  in \emph{International Workshop on Agents and P2P Computing}.\hskip 1em plus
  0.5em minus 0.4em\relax Springer, 2004, pp. 1--13.

\bibitem{ma2020temporal}
Y.~Ma, B.~Narayanaswamy, H.~Lin, and H.~Ding, ``Temporal-contextual
  recommendation in real-time,'' in \emph{Proceedings of the 26th ACM SIGKDD
  international conference on knowledge discovery \& data mining}, 2020, pp.
  2291--2299.

\bibitem{zhang2017improved}
B.~Zhang and B.~Yuan, ``Improved collaborative filtering recommendation
  algorithm of similarity measure,'' in \emph{AIP Conference Proceedings}, vol.
  1839, no.~1.\hskip 1em plus 0.5em minus 0.4em\relax AIP Publishing LLC, 2017,
  p. 020167.

\bibitem{cohen2009pearson}
I.~Cohen, Y.~Huang, J.~Chen, J.~Benesty, J.~Benesty, J.~Chen, Y.~Huang, and
  I.~Cohen, ``Pearson correlation coefficient,'' \emph{Noise reduction in
  speech processing}, pp. 1--4, 2009.

\bibitem{johnson1996cliques}
D.~S. Johnson and M.~A. Trick, \emph{Cliques, coloring, and satisfiability:
  second DIMACS implementation challenge, October 11-13, 1993}.\hskip 1em plus
  0.5em minus 0.4em\relax American Mathematical Soc., 1996, vol.~26.

\bibitem{bertin2011million}
T.~Bertin-Mahieux, D.~P. Ellis, B.~Whitman, and P.~Lamere, ``The million song
  dataset,'' 2011.

\bibitem{10.1145/2187980.2188222}
\BIBentryALTinterwordspacing
B.~McFee, T.~Bertin-Mahieux, D.~P. Ellis, and G.~R. Lanckriet, ``The million
  song dataset challenge,'' in \emph{Proceedings of the 21st International
  Conference on World Wide Web}, ser. WWW '12 Companion.\hskip 1em plus 0.5em
  minus 0.4em\relax New York, NY, USA: Association for Computing Machinery,
  2012, p. 909–916. [Online]. Available:
  \url{https://doi-org.tudelft.idm.oclc.org/10.1145/2187980.2188222}
\BIBentrySTDinterwordspacing

\bibitem{webber2010similarity}
W.~Webber, A.~Moffat, and J.~Zobel, ``A similarity measure for indefinite
  rankings,'' \emph{ACM Transactions on Information Systems (TOIS)}, vol.~28,
  no.~4, pp. 1--38, 2010.

\end{thebibliography}

\end{document}